\documentclass{elsarticle} 

\usepackage{times}
\usepackage{amssymb}
\usepackage{amsmath}
\usepackage{color}
\usepackage{graphicx}
\usepackage{eurosym}
\usepackage{amsthm}
\usepackage{relsize}
\usepackage{mathtools}
\usepackage{subfig}
\usepackage{float}
\usepackage{dsfont} 
\usepackage{multirow}
\usepackage{siunitx}
\usepackage{tabularx}
\usepackage{array}
\usepackage{natbib}

\usepackage{etoolbox}
\patchcmd{\MaketitleBox}{\footnotesize\itshape\elsaddress\par\vskip36pt}{\footnotesize\itshape\elsaddress\par\parbox[b][36pt]{\linewidth}{\vfill\hfill\textnormal{\today}\hfill\null\vfill}}{}{}%
\patchcmd{\pprintMaketitle}{\footnotesize\itshape\elsaddress\par\vskip36pt}{\footnotesize\itshape\elsaddress\par\parbox[b][36pt]{\linewidth}{\vfill\hfill\textnormal{\today}\hfill\null\vfill}}{}{}%

\begin{document}

\begin{frontmatter}
\title{The Gauss2++ Model -- 
A Comparison of Different Measure Change Specifications for a Consistent Risk Neutral and Real World Calibration}
\author[LMU,ROKOCO]{Christoph Berninger}
\author[ROKOCO]{Julian Pfeiffer}
\address[LMU]{Department of Statistics, LMU München}
\address[ROKOCO]{ROKOCO GmbH, Ludwig-Ganghofer-Str. 6, 82031 Grünwald}
\begin{abstract}
Especially in the insurance industry interest rate models play a crucial role e.g. to calculate the insurance company's liabilities, performance scenarios or risk measures. A prominant candidate is the \textit{2-Additive-Factor Gaussian Model (Gauss2++ model)} -- in a different representation also known as the \textit{2-Factor Hull-White model}. In this paper, we propose a framework to estimate the model such that it can be applied under the risk neutral and the real world measure in a consistent manner. We first show that any progressive and square-integrable function can be used to specify the change of measure without loosing the analytic tractability of e.g. zero-coupon bond prices in both worlds. 
We further propose two time dependent candidates, which are easy to calibrate: a step and a linear function. They represent two variants of our framework and distinguish between a short and a long term risk premium, which allows to regularize the interest rates in the long horizon. We apply both variants to historical data and show that they indeed produce realistic and much more stable long term interest rate forecast than the usage of a constant function. This stability over time would translate to performance scenarios of e.g. interest rate sensitive fonds and risk measures.
\end{abstract}
\end{frontmatter}

\newpage

\section{Introduction}
Two prominent approaches to model the term structure of interest rates are the classes of equilibrium and no-arbitrage models. Most equilibrium models concentrate on the dynamic of the short-rate -- the instantaneous interest rate – and derive interest rates with longer maturities from it. Prominent candidates of this model class include \citep{cox1985}, \citep{duffie1996} and \citep{vasicek1977}. No-arbitrage models focus on exactly fitting the term structure at a specific point in time to prevent arbitrage possibilities. Representatives of this class are introduced by \citep{heath1992} and \citep{hullwhite1990}. \\ 
Applications of these models often relate to pricing interest rate derivatives, which is the reason why they are directly defined under the risk neutral measure most of the time. A general form of a one-factor short-rate model under the risk neutral measure is, e.g., given by 
$$
dr(t) = \mu(t,r)dt + \sigma(t,r)dW(t),
$$
where $\mu$ and $\sigma$ are two functions, which can depend on time point $t$ and the short-rate $r$, and $W$ is a Brownian motion. A lot of advances in theoretic models and their estimation have been conducted in the last 30 years, but only in connection to pricing \citep{diebold2006}. Regarding these models little attention has been given to forecasting and risk management purposes \citep{diebold2006}. For these applications the corresponding model needs to be regarded under the real world measure. Under this measure the corresponding one factor short-rate model has the following dynamic
$$
dr(t) = \bigg[\mu(t,r) + \lambda(t,r) \sigma(t,r) \bigg]dt + \sigma(t,r)d\widetilde{W}(t),
$$
where $\lambda$ is the market price of risk and can also depend on $t$ and $r$. $\widetilde{W}$ is a Brownian motion under the real world measure. The exact functional choice for $\lambda$ completes the model specification under the real world measure. \citet{dai2000} as well as \citet{jong2000} use a fixed multiple of the model's variance for the market price of risk and investigate the in sample fit of specific short-rate models, but do not focus on forecasting. \citet{duffee2002} concludes that the class of term structure models analysed in \citep{dai2000} fail in forecasting.  He argues that a restriction for the market price of risk to be a fixed multiple of the variance reduces the flexibility of the model. \citet{hull2014} stress that the market price of risk for a model with few factors should be time dependent. This results not from an economic interpretation but from a modelling issue because of an insufficient number of factors \citep{hull2014}. They estimated the market price of risk based on historical 3-month and 6-month interest rates and came to a similar result as \citep{ahmad2006}, \citep{cox1999} and \citep{stanton1997}. But they argue that this value is only valid in the short horizon. Keeping this market price of risk constant could lead to extreme risk premiums and interest rates in the long horizon. \\
In this paper we tackle exactly this problem. Instead of assuming a constant, we assume a time-varying function for the market price of risk. In contrast to \citet{hull2014}, who estimate the market price of risk for each forecasting horizon individually, we propose two parametric functions. The step function is the easiest non-constant function, which allows to model a market price of risk valid in the short and one valid in the long horizon. The linear function assumes that the market price of risk in the short horizon converges linearly to a long-term level. With these simplified time dependent functions it is possible to account for the problem mentioned by \citet{hull2014} and the functions can still be easily estimated by historical data or calibrated in a forward looking manner to interest rate forecasts. 
\\
\\
The structure of the paper is as follows. In Section \ref{sec:Gauss2++Theory} we introduce the Gauss2++ model under the risk neutral and the real world measure in a very general framework. In Section \ref{sec:LLRRPF} we propose the constant function for comparison reason as well as the step and the linear function to specify the change of measure and explain how they can be estimated. All three variants of the Gauss2++ model are applied to data and backtested for the last 3 years in Section \ref{sec:Results}. In the final section the results are summarized and concluded.

\section{The Gauss2++ Model in the Risk Neutral and the Real World}
\label{sec:Gauss2++Theory}
Throughout this section a filtered probability space $(\Omega, \mathcal{F}, (\mathcal{F}_t)_{t \in [0,\mathcal{T}]}, \mathbb{M})$ is given, where $\mathbb{M}$ is either the risk neutral measure $\mathbb{Q}$ with respect to the bank account or the real world measure $\mathbb{P}$. $\mathcal{T}$ represents an appropriate modelling horizon. The bank account $(B(t))_{t \in [0,\mathcal{T}]}$ is given by
\begin{equation*}
dB(t) = r(t)B(t)dt, \qquad B(0) = 1.
\end{equation*}

\subsection{Short-Rate Models}
A challenge of modelling the yield curve is the multivariate setting as each interest rate with a specific maturity represents a dimension. Instead of modelling all maturities simultaniously, short-rate models just model the short-rate and derive interest rates with longer maturities via pricing zero-coupon bonds. Given the price of a zero-coupon bond, the corresponding interest rate can be calculated by
\begin{equation}
\label{eq:Zins}
r(t,T) = \frac{-ln(P(t,T))}{T-t},
\end{equation} 
where $r(t,T)$ and $P(t,T)$ represent the interest rate and the price of a zero-coupon bond at time $t$ and a maturity of $T$, respectively. 
\\
\\
For pricing zero-coupon bonds the financial mathematical method of risk neutral valuation can be applied. The risk neutral interest rates generated in this way can be used in a Monte Carlo simulation to price interest rate derivatives or bonds. This is the main application of short-rate models and the reason why they are often defined directly under the risk neutral measure. \\
The method of risk neutral valuation is a general concept in financial mathematics and uses the property, that price processes of any security in the market discounted by the bank account are martingales under $\mathbb{Q}$. Therefore, the risk neutral price of a zero-coupon bond at time point $t$ is obtained by
$$
\frac{P(t,T)}{B(t)} = E^{\mathbb{Q}}\left[\frac{P(T,T)}{B(T)} \bigg\vert \mathcal{F}_t \right].
$$
As the value of the bank account at time point $t$ is given by $B(t) = e^{\int_0^t r(s) ds}$ and the payoff of a zero-coupon bond is one amount of currency at time $T$ this leads to 
\begin{equation*}
P(t,T) = E^{\mathbb{Q}}\left[e^{-\int_t^T r(s) ds} \bigg\vert \mathcal{F}_t \right],
\end{equation*}
where $r(s)$ is the short-rate at time point $s$. If the distribution of $r(s)$ is known and such, that the conditional distribution of $e^{-\int_t^T r(s) ds}$ can be determined, zero-coupon bond prices of different maturities at different time points can be analytically calculated. From bond prices interest rates are available using (\ref{eq:Zins}), so that indeed the whole interest rate curve is characterized in terms of distributional properties of $r$. 
\\
\\
If one is not interested in pricing interest rate derivatives or bonds but in risk measures or performance scenarios, interest rates under the real world measure are needed. The challenge in the real world is that every financial product has a different drift in its process depending on its risk the (in general risk averse) investor wants to be compensated for. To get a martingale as in the risk neutral world such that we can use the conditional expectation to price a security in the market, we have to discount the price process with a cash flow, which is product specific and different from the risk neutral bank account. This cash flow is in general not known, which is the reason why one switches to the risk neutral world if interested in pricing and valuation. But by knowing the dynamics of the processes under the risk neutral measure and defining the change of measure, we implicitly define this cash flow for every security in the market and therefore we can calculate the price of a zero-coupon bond analogously with the conditional expectation
\begin{equation*}
\frac{P(t,T)}{X_{P(t,T)}(t)} = E^{\mathbb{P}}\left[ \frac{P(T,T)}{X_{P(t,T)}(T)} \bigg\vert \mathcal{F}_t\right],
\end{equation*}
where $X_{P(t,T)}(t)$ is the value of the cash flow at time point $t$, with which we have to discount $P(t,T)$ such that $\frac{P(t,T)}{X_{P(t,T)}(t)}$ is a martingale under $\mathbb{P}$. Note that we take the expectation under the real world measure $\mathbb{P}$. As $P(T,T)$ is one amount of currency the conditional expectation reduces to
\begin{equation}
\label{eq:CondExpRW}
P(t,T) = E^{\mathbb{P}}\left[\frac{X_{P(t,T)}(t)}{X_{P(t,T)}(T)} \bigg\vert \mathcal{F}_t \right].
\end{equation}
We will show in Section \ref{sec:changeOfMeasure} that if we define the change of measure in the Gauss2++ model in a specific way, $X_{P(t,T)}(t)$ can be easily extracted and a closed form solution for the price of a zero-coupon bond or interest rates can still be obtained.

\subsection{The Gauss2++ Model under the Risk Neutral Measure}
\label{sec:Gauss2++RiskNeutral}
Short-rate models differ in the underlying process for the short-rate. The Gauss2++ model assumes that the short-rate is given by a sum of two correlated normally distributed processes, $(x(t))_{t\in[0,\mathcal{T}]}$ and $(y(t))_{t\in[0,\mathcal{T}]}$, and a deterministic function $\varphi$, which is well defined on the time interval $[0,\mathcal{T}]$:
\begin{equation*}
r(t) = x(t) + y(t) + \varphi(t), \qquad r(0) = r_0,
\end{equation*} 
where $r_0$ is the short-rate at time point $0$. The processes $(x(t))_{t\in[0,\mathcal{T}]}$ and \linebreak $(y(t))_{t\in[0,\mathcal{T}]}$ satisfy under the risk neutral measure $\mathbb{Q}$ the following stochastic differential equations
\begin{align*}
dx(t) &= -ax(t)dt + \sigma dW^1(t), \qquad x(0) = 0,\\
dy(t) &= -by(t)dt + \eta dW^2(t), \qquad y(0) = 0,\\
\rho dt &= dW^1(t)dW^2(t),
\end{align*}
where $a$, $b$, $\sigma$, $\eta$ are non-negative constants and $-1 \leq \rho \leq 1$ is the instantaneous correlation between the two Brownian motions $W^1$ and $W^2$.\\
The short-rate is therefore normally distributed and it can be shown that $\int_t^T r(s) ds$ is also normally distributed with mean 
\begin{align*}
M(t,T) = \int_t^T \varphi(s) ds + B(a,t,T) x(t) + B(b,t,T) y(t)
\end{align*}
and variance \par
\begin{footnotesize}
\begin{align*}
V(t,T) &= \frac{\sigma^2}{a^2} \left[ (T-t) + \frac{2}{a}e^{-a(T-t)} - \frac{1}{2a}e^{-2a(T-t)} - \frac{3}{2a} \right] \notag \\
&+ \frac{\eta^2}{b^2} \left[ (T-t) + \frac{2}{b}e^{-b(T-t)} - \frac{1}{2b}e^{-2b(T-t)} - \frac{3}{2b} \right] \notag \\
&+ 2\rho \frac{\sigma \eta}{ab} \left[ (T-t) + \frac{e^{-a(T-t)}-1}{a} + \frac{e^{-b(T-t)}-1}{b} - \frac{e^{-(a+b)(T-t)}-1}{a+b} \right],
\end{align*}
\end{footnotesize} \par
\noindent where
\begin{equation*}
B(z,t,T) = \frac{1-e^{-z(T-t)}}{z}.
\end{equation*}
A derivation of the mean and the variance can be found in \citep{brigo2007}. \\
The expression $e^{-\int_t^T r(s) ds}$ is therefore log-normally distributed and the zero-coupon bond price $P(t,T)$, which is the conditional expectation of this expression, is given by
\begin{align}
\label{eq:BPFormula}
P(t,T) &= E^{\mathbb{Q}}\left[e^{-\int_t^T r(s) ds} \bigg\vert \mathcal{F}_t \right] \notag \\
&= e^{-M(t,T) + \frac{V(t,T)}{2}} \notag \\
&= e^{-\int_t^T \varphi(s) ds - B(a,t,T) x(t) - B(b,t,T) y(t) + \frac{1}{2} V(t,T)}.
\end{align}
With this closed form solution for the conditional expectation zero-\linebreak coupon bond prices under the risk neutral measure are readily defined and interest rates can be directly derived.
\\
\\
The financial market we actually model consists of a bank account and a set of zero-coupon bonds, $P(t,T)$, which differ in the maturity $T$. The dynamic of a zero-coupon bond price can be derived from the bond price formula in (\ref{eq:BPFormula}) by applying Ito's formula and is given by 
\begin{equation*}
\footnotesize
dP(t,T) = P(t,T) \bigg[r(t)dt - \sigma B(a,t,T) dW^1(t) - \eta B(b,t,T)dW^2(t)\bigg].
\end{equation*}
A detailed derivation can be found in \ref{proof:BPdynamicQ}. Note that all assets have the same drift as it is the case in the risk neutral world.

\subsection{The Gauss2++ Model under the Real World Measure}
\label{sec:changeOfMeasure}
To calculate performance scenarios and risk indicators the Gauss2++ model must be regarded under the real world measure $\mathbb{P}$. 

\subsubsection{The Change of Measure}
By specifying the Gauss2++ model under the risk neutral measure, we implicitly assume an arbitrage free market. Therefore, we can make the transition to a real world measure $\mathbb{P}$ by defining the change of measure according to Girsanov, who states that a progressive and square-integrable process \linebreak $(\boldsymbol{\Phi}(t))_{t\in[0,\mathcal{T}]}=\left(\Phi^1(t), \Phi^2(t),...,\Phi^d(t)\right)_{t\in[0,\mathcal{T}]}$ determines a new probability measure $\mathbb{P}$ such that if $(\boldsymbol{\widehat{W}}(t))_{t\in[0,\mathcal{T}]}$ is a standard $d$-dimensional $(\mathcal{F}_t)_{t \in [0,\mathcal{T}]}$-Brownian motion under $\mathbb{Q}$, then
$$
\boldsymbol{\breve{W}}(t) := \boldsymbol{\widehat{W}}(t) + \int_0^t \boldsymbol{\Phi}(s) ds
$$
defines a standard $d$-dimensional $(\mathcal{F}_t)_{t \in [0,\mathcal{T}]}$-Brownian motion under $\mathbb{P}$ \citep{girsanov1960}.
We can choose any $\boldsymbol{\Phi}$, which fullfills the conditions in the Girsanov theorem, to specify the change of measure. 
\\
\\
The Gauss2++ model is a two-factor model and $\boldsymbol{\Phi}$ is therefore 2-dimensional. Its components can be interpreted as the market price of risk for each factor in the model. We will represent $\boldsymbol{\Phi}$ as follows to simplify calculations 
\begin{align}
\label{eq:PhiDefinition}
\boldsymbol{\Phi}(t) = \left( \begin{array}{c} \Phi^1(t) \\ \Phi^2(t) \end{array} \right) = \left( \begin{array}{c} -\frac{ad_x(t)}{\sigma} \\ -\frac{bd_y(t)}{\eta \sqrt{1-\rho^2}} + \frac{\rho ad_x(t)}{\sigma \sqrt{1-\rho^2}} \end{array} \right).
\end{align}
Note that we have not restricted the set of functions by this representation. The conditions for the Girsanov theorem translate directly to the functions $d_x(t)$ and $d_y(t)$. In the following we will specify the change of measure via $d_x(t)$ and $d_y(t)$. 
An appropriate interpretation of these functions will be given in Section \ref{sec:DynamicsUnderP}.

\subsubsection{The dynamics under the real world measure $\boldsymbol{\mathbb{P}}$}
 \label{sec:DynamicsUnderP}
With the representation of $\boldsymbol{\Phi}$ as in (\ref{eq:PhiDefinition}) the dynamics of the processes $x$ and $y$ in the Gauss2++ model change according to Girsanov to
\begin{alignat}{2}
\label{eq:xUnderP}
dx(t) &= a(d_x(t)-x(t))dt + \sigma d\widetilde{W}^1(t), \qquad x(0) &&= 0,\\
\label{eq:yUnderP}
dy(t) &= b(d_y(t)-y(t))dt + \eta d\widetilde{W}^2(t), \qquad y(0) &&= 0,
\end{alignat}
where $\widetilde{W}^1$ and $\widetilde{W}^2$ are two correlated Brownian motions under $\mathbb{P}$. The derivation can be found in \ref{proof:xyUnderP}. We observe that $x$ and $y$ are still Ornstein-Uhlenbeck processes with the solutions
\begin{align}
\label{eq:RPFkt1}
x(t) &= \int_0^t e^{-a(t-u)}ad_x(u)du + \sigma \int_0^t e^{-a(t-u)} d\widetilde{W}(u), \\
\label{eq:RPFkt2}
y(t) &= \int_0^t e^{-b(t-u)}bd_y(u)du + \eta \int_0^t e^{-b(t-u)} d\widetilde{W}(u).
\end{align}
The mean reversion level of each process at time point $t$ amounts to $d_x(t)$ and $d_y(t)$, respectively. Recall that the sum of $x(t)$ and $y(t)$ and a deterministic function $\varphi(t)$ under the risk neutral measure adds up to the instantaneous return rate $r(t)$ of a risk free investment. Changing the measure changes the mean reversion level at time point $t$ from $0$ to $d_x(t)$ for the process $x$ and to $d_y(t)$ for the process $y$. Therefore, {$d_x(t)+d_y(t)$} can be interpreted as the local long run risk premium of the short-rate -- the amount, which is added in the real world to the risk neutral short-rate in the long run, if \linebreak {$d_x(t)+d_y(t)$} would stay constant over time. If this amount is negative, future bond prices increase in expectation compared to the risk neutral world and a risk averse investor, therefore, gets compensated for the risk of investing in a risky bond. This means in contrast to equity prices, in a market where investors are risk averse, future interest rates tend to be lower in the real world than in the risk neutral world \citep{hull2014}. Therefore, $d_x(t)$ and $d_y(t)$ can be interpreted as the local long run risk premium the corresponding risk factor is mean reverting to at time point $t$. \\
In the following we will specify the change of measure by these two functions instead of the market prices of risk. The market price of risk of each risk factor is then directly defined by these two functions.
\begin{align*}
&\text{Market price of risk of risk factor 1:} -\frac{ad_x(t)}{\sigma} \\
&\text{Market price of risk of risk factor 2:} -\frac{bd_y(t)}{\eta \sqrt{1-\rho^2}} + \frac{\rho ad_x(t)}{\sigma \sqrt{1-\rho^2}} .
\end{align*}
If we assume a step or a piecewise linear function for $d_x(t)$ and $d_y(t)$ the functional form of the individual market prices of risk are the same.
\\
\\
%
\noindent The dynamics of a zero-coupon bond with maturity $T$ under $\mathbb{P}$ has the following form
\begin{align}
\label{eq:BPdynamicRW}
\footnotesize
dP(t,T) = &P(t,T)\left[r(t) - B(a,t,T)ad_x(t) - B(b,t,T)bd_y(t)\right]dt \notag \\
&- P(t,T)B(a,t,T)\sigma d\widetilde{W}^1(t) - P(t,T)B(b,t,T)\eta d\widetilde{W}^2(t)
\end{align}
The derivation can be found in \ref{proof:BPdynamicP}. 

\subsubsection{The Bond Price Formula under The Real World Measure}
To calculate the price of a zero-coupon bond under the real world measure with the conditional expectation in (\ref{eq:CondExpRW}), the cash flow $X_{P(t,T)}$, with which we have to discount the zero-coupon bond such that the discounted price process is a martingale under $\mathbb{P}$, needs to be determined. The dynamic of $X_{P(t,T)}$ coincides with the deterministic part of the zero-coupon bond price dynamic in (\ref{eq:BPdynamicRW}) and is therefore specified by the change of measure: \par
\begin{footnotesize}
\begin{equation*}
dX_{P(t,T)}(t) = X_{P(t,T)}(t)\left[r(t) - B(a,t,T)ad_x(t) - B(b,t,T)bd_y(t)\right]dt, \qquad X_{P(t,T)}(0) = 1.
\end{equation*}
\end{footnotesize} \par
\noindent A short proof can be found in \ref{proof:DiskRateP}. The solution of this dynamic is given by
$$
X_{P(t,T)}(t) = e^{\int_0^t \left(r(u) - B(a,u,T) a d_x(u) - B(b,u,T) bd_y(u) \right) du}.
$$
As $\frac{P(t,T)}{X_{P(t,T)}(t)}$ is a martingale we can use the conditional expectation in (\ref{eq:CondExpRW}) to calculate the price of a zero-coupon bond at time point $t$:
\begin{equation*}
P(t,T) = E^{\mathbb{P}} \left[\frac{X_{P(t,T)}(t)}{X_{P(t,T)}(T)} \bigg\vert \mathcal{F}_t \right].
\end{equation*}
The ratio in the expectation amounts to
\begin{align*}
\frac{X_{P(t,T)}(t)}{X_{P(t,T)}(T)} = e^{-\int_t^T \left(r(u) - B(a,u,T) a d_x(u) - B(b,u,T) bd_y(u) \right) du}.
\end{align*}
To determine the distribution of this ratio, we first derive the distribution of the integral in the exponent, i.e.,
\begin{align*}
I(t,T) \coloneqq \int_t^T \left(r(u) - B(a,u,T) a d_x(u) - B(b,u,T) bd_y(u) \right) du.
\end{align*}
It can be shown that $I(t,T)$ is normally distributed with mean
\begin{align}
\label{eq:RWmean}
M(t,T) = \int_t^T \varphi(u)du + \frac{1-e^{-a(T-t)}}{a} x(t) + \frac{1-e^{-b(T-t)}}{b} y(t)
\end{align}
and variance \par
\begin{footnotesize}
\begin{align}
\label{eq:RWvariance}
V(t,T) &= \frac{\sigma^2}{a^2} \left[ (T-t) + \frac{2}{a}e^{-a(T-t)} - \frac{1}{2a}e^{-2a(T-t)} - \frac{3}{2a} \right] \notag \\
&+ \frac{\eta^2}{b^2} \left[ (T-t) + \frac{2}{b}e^{-b(T-t)} - \frac{1}{2b}e^{-2b(T-t)} - \frac{3}{2b} \right] \notag \\
&+ 2\rho \frac{\sigma \eta}{ab} \left[ (T-t) + \frac{e^{-a(T-t)}-1}{a} + \frac{e^{-b(T-t)}-1}{b} - \frac{e^{-(a+b)(T-t)}-1}{a+b} \right].
\end{align}
\end{footnotesize} \par
\noindent The variance is the same as in the risk neutral world as the change of measure does not influence the variance of the processes. Note that also the mean has the same form as in the risk neutral case as the terms $B(a,u,T)ad_x(u)$ and $B(b,u,T)bd_y(u)$ in $I(t,T)$ cancel out in the calculations. The derivations can be found in \ref{proof:BPFormulaP}.
\\
\\
The expression $e^{-I(t,T)}$ is therefore log-normally distributed and the zero-coupon bond price under $\mathbb{P}$ is given by
\begin{align*}
P(t,T) &= E^{\mathbb{P}}\left[ e^{-\int_t^T r(u) - B(a,u,T) a d_x(u) - B(b,u,T) bd_y(u) du} \mid \mathcal{F}_t\right] \\
&= e^{-M(t,T) + \frac{1}{2} V(t,T)} \\
&= e^{-\int_t^T \varphi(u)du - \frac{1-e^{-a(T-t)}}{a} x(t) - \frac{1-e^{-b(T-t)}}{b} y(t) + \frac{1}{2}V(t,T)}.
\end{align*}
The bond price formula stays, therefore, exactly the same as in the risk neutral case. The only difference is, that $x(t)$ and $y(t)$ are now the values at time point $t$ of the corresponding processes under the real world measure $\mathbb{P}$.

\section{Local Long Run Risk Premium Functions -- Specification and Calibration}
\label{sec:LLRRPF}
\begin{figure}[H]%
  \centering
  \subfloat[][]{\includegraphics[width=0.33\linewidth]{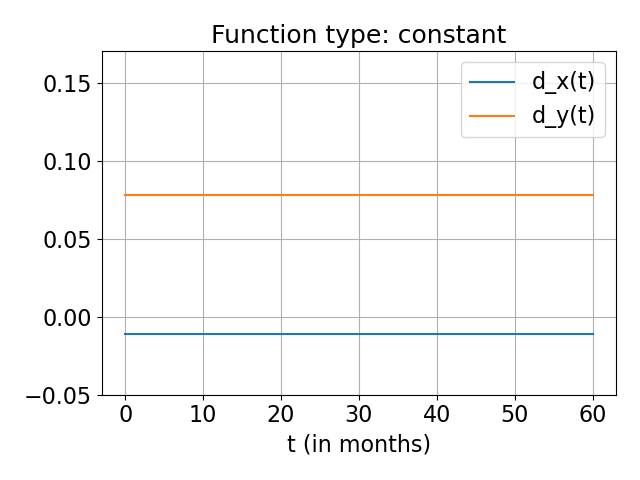}}%
  \subfloat[][]{\includegraphics[width=0.33\linewidth]{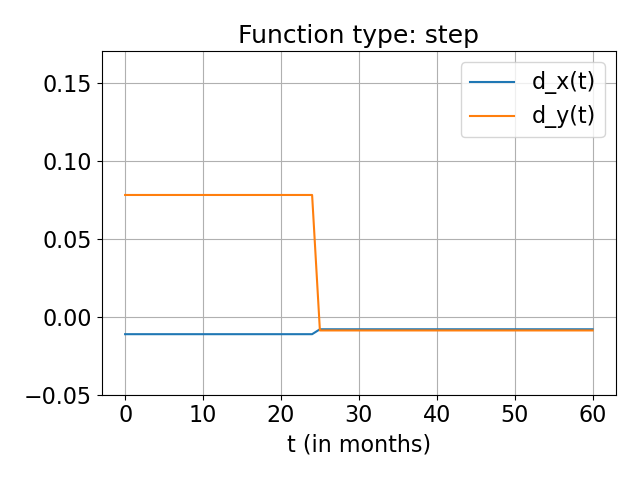}}%
  \subfloat[][]{\includegraphics[width=0.33\linewidth]{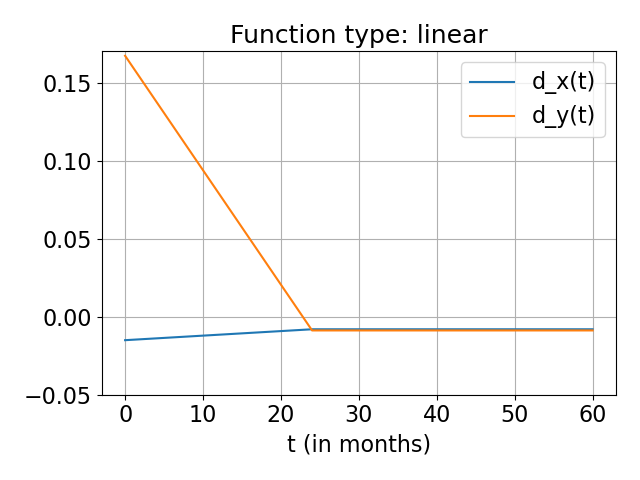}}%
  \caption{Local long run risk premium functions}%
  \label{InstRPFkt}
\end{figure}

In the following three different types of functions for $d_x(t)$ and $d_y(t)$ are introduced: the constant, the step and the linear function. Following the interpretation in Section \ref{sec:DynamicsUnderP} these functions represent the long run risk premium for each risk factor at a specific time point $t$ in the Gauss2++ model. The functional equations of the three types are 
\begin{table}[H]
\begin{tabular}{l l l}

Constant:	& $d_x(t) = d_x$ \\
			& $d_y(t) = d_y$ \\
			\\
Step:		& $d_x(t) = \mathds{1}_{t\leq\tau}d_x + \mathds{1}_{t>\tau}l_x $ \\
			& $d_y(t) = \mathds{1}_{t\leq\tau}d_y + \mathds{1}_{t>\tau}l_y $ \\ 
			\\
Linear: 	& $d_x(t) = \mathds{1}_{t\leq\tau}(1-m_xt)d_x + \mathds{1}_{t>\tau}l_x $ \\
			& $d_y(t) = \mathds{1}_{t\leq\tau}(1-m_yt)d_y + \mathds{1}_{t>\tau}l_y $\\
\end{tabular}
\end{table}
\noindent where $d_x$, $l_x$, $m_x$ and $d_y$, $l_y$, $m_y$ are real valued constants and $\mathds{1}_{A}$ represents the indicator function of a subset $A$.\\
The constant function assumes that the local long run risk premium is constant for the whole modelling horizon. The latter two functions distinguish between a local long run risk premium valid in the short and in the long horizon, seperated at time point $\tau$. As mentioned in Section \ref{sec:DynamicsUnderP} the same holds for the market price of risk, respectively. \citet{hull2014} argue that a time varying market price of risk is necessary to account for unobserved risk factors and to prevent unrealistic interest rate forecasts in the long horizon. They therefore estimate an individual market price of risk for each forecasting horizon. We use a more parsimonious function with regard to the number of parameters. The step function we propose is the simplest time varying function that expects that the local long run risk premium differs in the short and the long horizon but is still constant in each period. The linear function implements the property that the local long run risk premium in the short horizon approaches the long term level linearly. The simplicity of these functions allows a straight forward calibration to interest rate forecasts.
\\
\\
Because of the distributional properties of the Gauss2++ model the expected values for interest rates under the real world measure $\mathbb{P}$ for any future time point can be calculated:
\begin{align}
\label{eq:IRExp}
E^{\mathbb{P}}[r(t,T)] &= E^{\mathbb{Q}}[r(t, T)] + \frac{B(a,t,T)}{T-t}  RP_x(t) + \frac{B(b,t,T)}{T-t} RP_y(t), 
\end{align}
where $RP_x(t)$ and $RP_y(t)$ represent the actual risk premium of the short-rate at time point $t$ for each risk factor and are given by the first integral in (\ref{eq:RPFkt1}) and (\ref{eq:RPFkt2})
\begin{align*}
RP_x(t) &\coloneqq \int_0^t e^{-a(t-u)}ad_x(u)du, \\
RP_y(t) &\coloneqq \int_0^t e^{-b(t-u)}bd_y(u)du.
\end{align*} 
For the constant, the step and the linear function these integrals can be easily calculated. To get the risk premium for longer maturities the functions $RP_x(t)$ and $RP_y(t)$ are weighted by a loading function, which accounts for the different riskiness of the corresponding zero-coupon bonds
\begin{align*}
\frac{B(a,t,T)}{T-t} \qquad  \text{and} \qquad \frac{B(b,t,T)}{T-t}.
\end{align*}
To calibrate the local long run risk premium functions, $d_x(t)$ and $d_y(t)$, the parameters of the functions are chosen in such a way that the model meets specific interest rate forecasts in expectation. For the constant type two interest rate forecasts are needed. For the other two types four interest rate forecasts are necessary -- two short term and two long term forecasts. The time parameter $\tau$, which determines the separation between the short and the long term local long run risk premium must lie between the forecasting horizons of the two short and the two long term forecasts. 
\\
\\
In Figure \ref{InstRPFkt} the three types of local long run risk premium functions have been exemplary calibrated.
$\tau$ has been set to $24$ months, which is the forecasting horizon of the short term interest rate forecasts.
\noindent In the following subsections the calibration procedures for all three types of local long run risk premium functions, which are applied in this paper, are described.

\subsection{The Constant Function}
\label{sec:CalibrateConstantdxdy}
The constant functions represented in Figure \ref{InstRPFkt} (a) implement a constant local long run risk premium for the whole modelling horizon, which can amount to up to $40$ years for actual applications in the insurance industry, e.g., to classify certified pension contracts into risk classes. The absolute risk premiums, $RP_x(t)$ and $RP_y(t)$, are given by:
\begin{align*}
RP_x(t) &= (1-e^{-at})d_x, \\
RP_y(t) &= (1-e^{-bt})d_y.
\end{align*}
Note that if $t \rightarrow \infty$, $RP_x(t)$ and $RP_y(t)$ indeed converge to $d_x$ and $d_y$, the long run risk premiums, respectively. To calibrate the parameters of the constant functions two interest rate forecasts, $ \hat{r}(t_1, T_1) $ and $\hat{r}(t_2, T_2)$, are used. Plugging the absolute risk premium functions, $RP_x(t)$ and $RP_y(t)$, into (\ref{eq:IRExp}) and setting the expectations equal to the interest rate forecasts results in the following two equations\par
\begin{footnotesize}
\begin{center}
\begin{tabular}{c l}
(I)& $ \hat{r}(t_1, T_1) \overset{!}{=} E^Q[r(t_1, T_1)] +  \frac{B(a,t_1,T_1)}{(T_1-t_1)} (1-e^{-at_1})d_x + \frac{B(b,t_1,T_1)}{(T_1-t_1)} (1-e^{-bt_1})d_y $,\\
(II)& $ \hat{r}(t_2, T_2) \overset{!}{=} E^Q[r(t_2, T_2)] +  \frac{B(a,t_2,T_2)}{(T_2-t_2)} (1-e^{-at_2})d_x + \frac{B(b,t_2,T_2)}{(T_2-t_2)} (1-e^{-bt_2})d_y $.
\end{tabular}
\end{center}
\end{footnotesize} \par
\noindent As the expectations are linear functions in $d_x$ and $d_y$, the two parameters can be easily determined.
\\
\\
The constant function for the local long run risk premium in the Gauss2++ model and this calibration procedure is a standard approach in the insurance industry. As the values for $d_x$ and $d_y$ determine the risk premium for the whole modelling horizon, their calibration is crucial for the model's interest rate distribution. Especially if the interest rate forecasts used for the calibration have a short forecasting horizon, the resulting distribution in the long horizon is very sensitive to these forecasts. For example if the interest rate forecasts and the forward rates -- calculated from the current yield curve -- are very different, to reach the forecasts a huge risk premium is necessary, which might be valid in the short horizon, but produces extreme interest rates in the long horizon. The next two functions account for this problem by representing a time varying local long run risk premium.

\subsection{The Step Function}
\label{sec:CalibrateStepdxdy}
The step functions represented in Figure \ref{InstRPFkt} (b) take the same value as the corresponding constant function up to time $\tau$ as the same interest rate forecasts have been used for the short horizon, but then they jump to a different level to account for the risk premium in the long horizon. Similar to the constant function the absolute risk premium functions can easily be calculated and amount to
\begin{align*}
RP_x(t) &= \left(e^{-a(t-\min(t,\tau))}-e^{-at}\right)d_x + \left(1-e^{-a (t - \min(t,\tau))}\right)l_x, \\
RP_y(t) &= \left(e^{-b(t-\min(t,\tau))}-e^{-bt}\right)d_y + \left(1-e^{-b (t- \min(t,\tau))}\right)l_y.
\end{align*}
Note that if $t \rightarrow \infty$, $RP_x(t)$ and $RP_y(t)$ now converge to $l_x$ and $l_y$, respectively. To calibrate the four parameters of the step function two short term and two long term interest rate forecasts are used resulting in the following equations:\par
\begin{footnotesize}
\begin{center}
\begin{tabular}{c l}
(I)& $ \hat{r}(t_1, T_1) \overset{!}{=} E^Q[r(t_1, T_1)] +  \frac{B(a,t_1,T_1)}{(T_1-t_1)} RP_x(t_1) + \frac{B(b,t_1,T_1)}{(T_1-t_1)} RP_y(t_1) $,\\
(II)& $ \hat{r}(t_2, T_2) \overset{!}{=} E^Q[r(t_2, T_2)] +  \frac{B(a,t_2,T_2)}{(T_2-t_2)} RP_x(t_2) +\frac{B(b,t_2,T_2)}{(T_2-t_2)} RP_y(t_2) $, \\
(III)& $ \hat{r}(t_3, T_3) \overset{!}{=} E^Q[r(t_3, T_3)] +  \frac{B(a,t_3,T_3)}{(T_3-t_3)} RP_x(t_3) + \frac{B(b,t_3,T_3)}{(T_3-t_3)} RP_y(t_3) $,\\
(IV)& $ \hat{r}(t_4, T_4) \overset{!}{=} E^Q[r(t_4, T_4)] +  \frac{B(a,t_4,T_4)}{(T_4-t_4)} RP_x(t_4) + \frac{B(b,t_4,T_4)}{(T_4-t_4)} RP_y(t_4) $,
\end{tabular}
\end{center}
\end{footnotesize} \par
\noindent where $t_1 \leq t_2 < t_3 \leq t_4 $. $\tau$ must lie between $t_2$ and $t_3$, i.e. $t_2 \leq \tau < t_3$. \\
Instead of interest rate forecasts direct forecasts of the absolute risk premium of the short-rate can be used. This approach is applied by \citet{hull2014}, who estimate risk premiums for each forecasting horizon from historical data, but they also scale their result to a long term short-rate forecast.

\subsection{The Linear Function}
\label{sec:CalibrateLineardxdy}
The linear functions represented in Figure \ref{InstRPFkt} (c) avoid the sudden jump as it is the case in the step functions and converge in the short term linearly to a long term level. The absolute risk premiums at time point $t$ can be calculated as before and amount to \par
\begin{footnotesize}
\begin{align*}
RP_x(t) &= \left ( \left(e^{-a(t-\min(t,\tau))}-e^{-at}\right)\left(1+\frac{m_x}{a}\right) + e^{-a (t-\min(t,\tau))}m_x\min(t,\tau)\right) d_x \\
&+ \left(1-e^{-a (t-\min(t,\tau))}\right)l_x, \\
RP_y(t) &= \left(\left(e^{-b(t-\min(t,\tau))}-e^{-bt}\right)\left(1+\frac{m_y}{b}\right) + e^{-b (t-\min(t,\tau))}m_y\min(t,\tau)\right)d_y \\
&+ \left(1-e^{-b (t-\min(t,\tau))}\right)l_y.
\end{align*}
\end{footnotesize} \par
\noindent Note again that if $t \rightarrow \infty$, $RP_x(t)$ and $RP_y(t)$ converge to $l_x$ and $l_y$, the long term risk premiums, respectively. To calibrate $d_x$, $l_x$, $d_y$ and $l_y$ four interest rate forecasts as for the step function are used. By imposing that the absolute risk premium functions, $RP_x(t)$ and $RP_y(t)$, are differentiable at the forecasting horizon $\tau$ to prevent a kink in the absolute risk premium function, two further conditions are incorporated to specify $m_x$ and $m_y$:
\begin{center}
\begin{tabular}{c l}
(V)& $ RP_x'(t) \big|_{t = \tau-} = RP_x'(t) \big|_{t = \tau +} $,\\
(VI)& $RP_y'(t) \big|_{t = \tau-} = RP_y'(t) \big|_{t = \tau +}$.
\end{tabular}
\end{center}
Solving the equations for $m_x$ and $m_y$ leads to the following closed form solutions reducing the number of free parameters to four:
\begin{align*}
m_x &= \frac{d_x - l_x}{d_x  \tau}, \\
m_y &= \frac{d_y - l_y}{d_y  \tau}.
\end{align*}
Note that with this condition the same number of interest rate forecasts as for the step function are needed to calibrate $d_x(t)$ and $d_y(t)$.

\section{Results}
\label{sec:Results}
In this Section the calibration results of three variants of our framework for the Gauss2++ model are presented. The variants differ in the assumption about the local long run risk premium functions, which determine the change from the risk neutral to the real world measure. Variant 1 assumes a constant, variant 2 a step and variant 3 a linear local long run risk premium function for the risk factors. In the first Subsection the three variants of the Gauss2++ model are compared if calibrated at the same valuation date. In Subsection \ref{sec:Backtest} we show with a backtest over the last three years that variant 2 and 3 produce much more stable interest rate scenarios for the long forecasting horizon over this time period. This stability would transfer to performance scenarios and risk measures of e.g. an interest rate sensitive fonds.

\subsection{Calibration at One Valuation Date}
\label{sec:OneValuationDate}
The calibration process of the Gauss2++ model can be split into two steps. In the first step the model is calibrated under the risk neutral measure. This step does not depend on the choice of the local long run risk premium function and is therefore the same for all modelling cases. In the second step the change of measure is calibrated. The choice of the local long run risk premium function plays an important role and leads to different interest rate scenarios, performance measures and risk indicators.
\\
\\
To calibrate the model at a specific valuation date under the risk neutral measure the term structure of interest rate swaps and swaption volatilities at this date are used. The Gauss2++ model presumes a specific dynamic for the short-rate and with it for interest rates with longer maturities. The parameters of the model are chosen in such a way, that the current term structure is met in expectation and that the model prices of the swaptions coincide with the market prices. In this way market consistency of the model is ensured. As $\varphi$ is a deterministic function of time, a perfect fit in expectation to the current term structure of interest rates can be achieved, i.e. the function $\varphi$ is implicitly given by the current interest rate curve. Later in the modelling process we use the term structure of german government bond yields with the assumption that the dynamic of this term structure is the same as for the term structure of interest rate swaps. For the calibration of the five parameters $a,b,\sigma,\eta$ and $\rho$ the downhill simplex algorithm is used to find the parameter set, which replicates the market swaption prices best. Table \ref{tbl:ParameterRN31122019} shows the results of a calibration at the 31.12.2019. We use swaptions with a maturity and tenor combination of $\{5,7,10,12,15,20\}$ x $\{5,7,10,12,15,20\}$, i.e. in total $36$ swaption prices.
\begin{table}
\newcolumntype{M}[1]{>{\centering\arraybackslash}b{#1}}
\begin{center}
\begin{tabularx}{\textwidth}{M{2.25cm}M{2.25cm}M{2.25cm}M{2.25cm}M{2.25cm}}
$ a $    & $ b $    & $\sigma$ & $ \eta $ & $ \rho $ \\ \hline \hline
\\
$ 0.2997 $ & $ 0.0407 $ & $ 0.0114 $  & $ 0.0114 $ & $ -0.9998 $ \\
\hline
\end{tabularx}
\caption{Parameters of the Gauss2++ model calibrated at 31.12.2019}
\label{tbl:ParameterRN31122019}
\end{center}
\end{table}
\noindent These parameters together with the current interest rate curve determine the dynamics of the Gauss2++ model under the risk neutral measure.
\\
\\
In the second step the local long run risk premium functions, which determine the change of measure, are calibrated to interest rate forecasts as described in Section \ref{sec:CalibrateConstantdxdy}-\ref{sec:CalibrateLineardxdy}. For the short term interest rate forecasts we use forecasts published by the OECD for a 3-month and a 10-year interest rate. The latest forecasts regarding the 31.12.2019 for the longest horizon, which is the fourth quarter of 2021, amount to $-0.4\%$ and $0.4\%$, respectively\footnote{https://stats.oecd.org}. For the long term interest rate forecasts, which are needed to calibrate the step and the linear function, we take the average of monthly 3-month and 10-year interest rates over the last 15 years also published by the OECD. This is a valid approach if interest rates follow a stationary process, because in this case historical data can be considered as a random sample from the corresponding interest rate distribution. \citet{hull2014} point out that this approach is questionable if monetary and fiscal policies are expected to be materially different from those in the past. Nevertheless any other model based on historical data would be questionable and the user of the model can alternatively provide personal estimates or an expert judgment. The historical average amounts to $1.08\%$ for the 3-month and $1.84\%$ for the 10-year interest rate and as we assume these forecasts to be a long run average we set the forecasting horizon to $40$ years -- the modelling horizon. We further set $\tau$ to 24 months, which is the forecasting horizon of the short term OECD forecasts. 
\\
\\
Table \ref{tbl:ParameterRW31122019} shows the calibration results for the three local long run risk premium function types. 

\begin{figure}[H]%
  \centering
  \subfloat[][]{\includegraphics[width=0.33\linewidth]{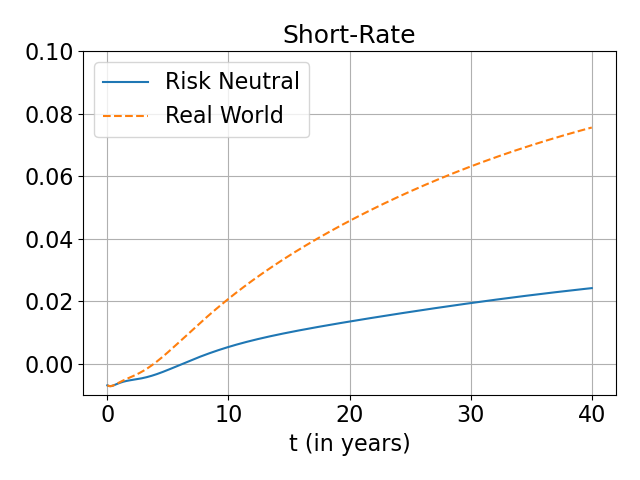}}%
  \subfloat[][]{\includegraphics[width=0.33\linewidth]{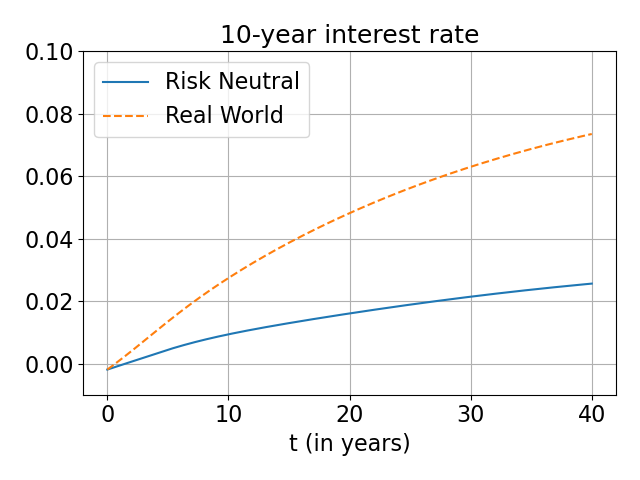}}%
  \subfloat[][]{\includegraphics[width=0.33\linewidth]{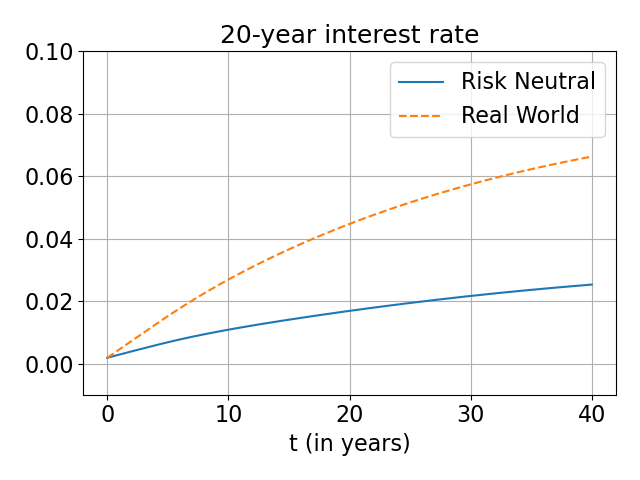}}%
  \caption{Constant Function}%
  \label{ConstRPFkt}
\end{figure}

\begin{figure}[H]%
  \centering
  \subfloat[][]{\includegraphics[width=0.33\linewidth]{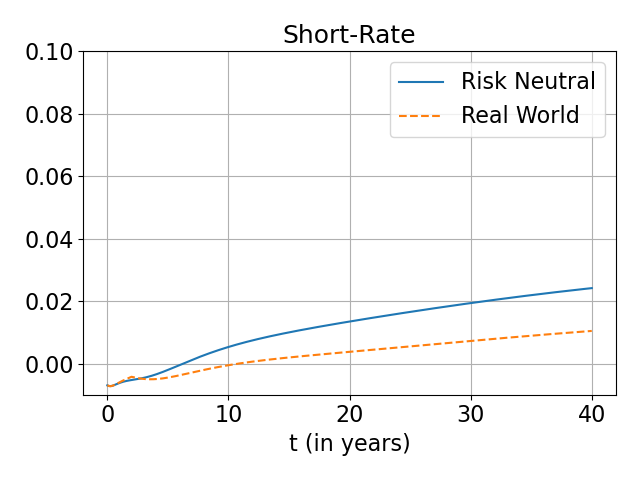}}%
  \subfloat[][]{\includegraphics[width=0.33\linewidth]{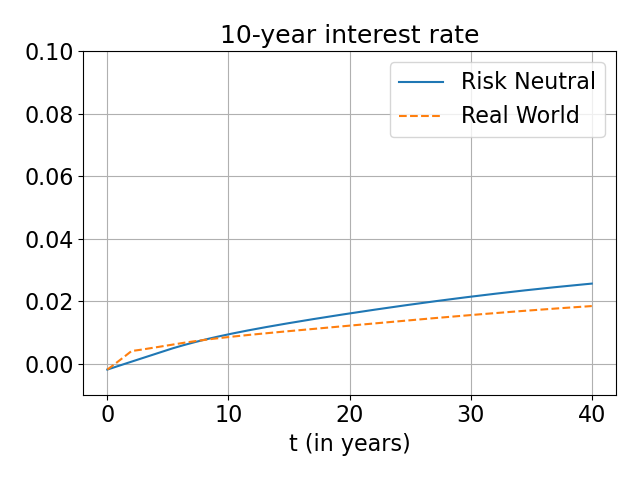}}%
  \subfloat[][]{\includegraphics[width=0.33\linewidth]{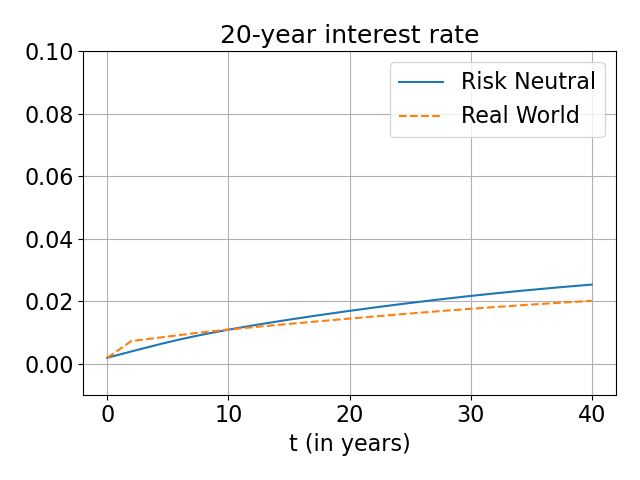}}%
  \caption{Step Function}%
  \label{StepRPFkt}
\end{figure}

\begin{figure}[H]%
  \centering
  \subfloat[][]{\includegraphics[width=0.33\linewidth]{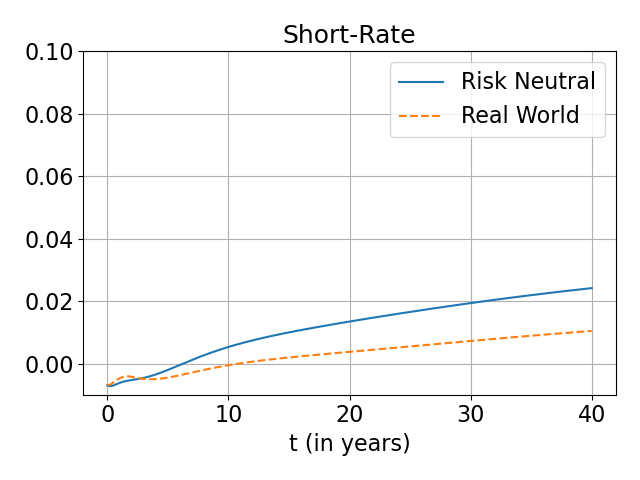}}%
  \subfloat[][]{\includegraphics[width=0.33\linewidth]{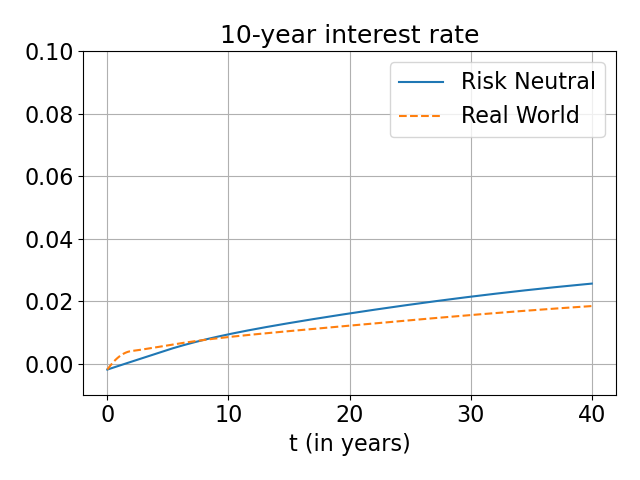}}%
  \subfloat[][]{\includegraphics[width=0.33\linewidth]{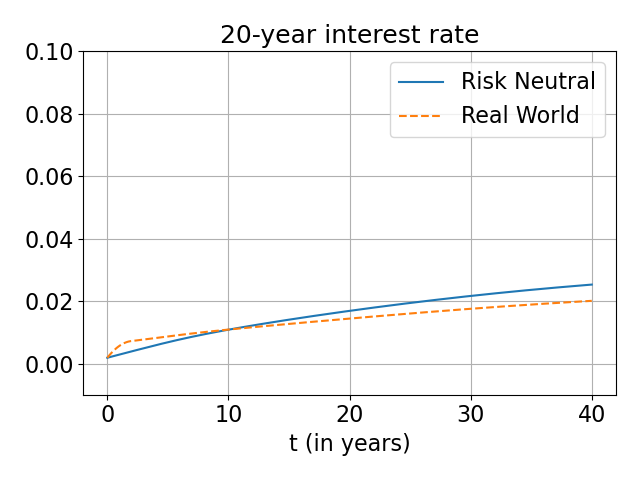}}%
  \caption{Linear Function}%
  \label{LinearRPFkt}
\end{figure}

\begin{figure}%
  \centering
\includegraphics[width=0.6\linewidth]{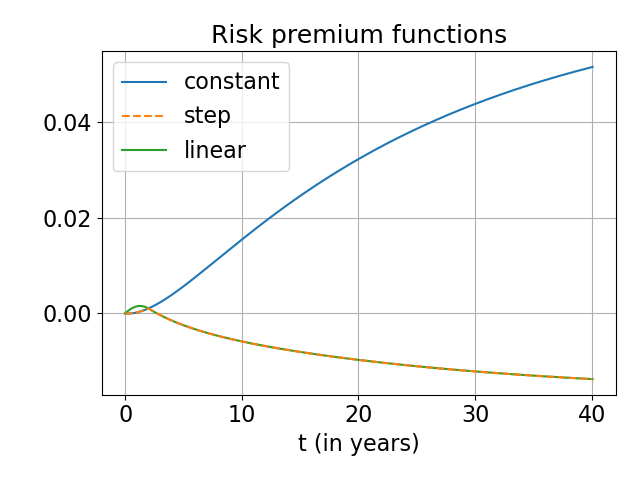}%
 \caption{Absolute risk premium function for the variants of the Gauss2++ model}
\label{fig:ThreeRPFunctions}
\end{figure}

\begin{table}[H]
\newcolumntype{M}[1]{>{\centering\arraybackslash}b{#1}}
\begin{center}
\begin{tabularx}{\textwidth}{p{3cm}M{1.85cm}M{1.85cm}M{1.85cm}M{1.85cm}}
 & $ d_x  $ & $ d_y $ & $ l_x $ & $ l_y $ \\\hline \hline
 \\
 \textit{Constant Function} & $ -0.0112 $ & $ 0.0779 $ & & \\
 \\
 \textit{Step Function} &$ -0.0112 $ & $ 0.0779 $ & $ -0.0081 $ & $ -0.0088 $ \\ 
 \\
 \textit{Linear Function} &$ -0.0151 $ & $ 0.1672 $ & $ -0.0081 $ & $ -0.0088 $ \\ \hline
\end{tabularx}
\caption{Parameters of the local long run risk premium functions}
\label{tbl:ParameterRW31122019}
\end{center}
\end{table}
\noindent The values of $d_x$ and $d_y$ coincide for the constant and the step function as the same interest rate forecasts have been used in the calibration process. But in contrast to the step function, which takes the values of $l_x$ and $l_y$ after $24$ months, the constant function stays constant for the whole modelling horizon. It also appears that the step and the linear function take the same values for $l_x$ and $l_y$. But there is a slight difference as their functional forms differ in the first two years, which influences the absolute risk premium in future time points. This influence decreases in time, such that the difference is negligible as we calibrated $l_x$ and $l_y$ to forecasts with an forecasting horizon of $40$ years.
\\
\\
Figure \ref{ConstRPFkt}-\ref{LinearRPFkt} visualize for the three calibrated variants of the Gauss2++ model the development of the expectation of the short-rate, the $10$-year and the $20$-year interest rate for forecasting horizons of up to  40 years. The solid line represents the expectation under the risk neutral measure, the dashed line shows the expected values under the real world measure.
\noindent For the variant of the Gauss2++ model, which uses the constant function as the local long run risk premium function, the expected real world interest rates lie above the risk neutral expectation. This means, that a risk seeking behaviour of the investors is assumed for the whole modelling period, because an investor accepts a lower expected return for a corresponding bond if the interest rates are expected to be higher in the real world compared to the risk neutral world. \citet{ahmad2006} show that there have been time periods where investors seem to have historically behaved in this way. But in general investors are assumed to be risk averse and therefore interest rates should be lower in the real world than in the risk neutral world, which is an opposite behaviour to equity prices \citep{hull2014}. For the other two variants of the Gauss2++ model the expected real world interest rates lie also above the risk neutral interest rates in the short horizon but below in the long horizon. This assumption of risk seeking behaviour in the short horizon stems from the quite high forecasts of the OECD for the short horizon, but it might be valid in the current market situation. In contrast to the constant case, which keeps this risk seeking behaviour assumption for the whole modelling horizon, in the long run the other two variants of the Gauss2++ model assume in this calibration a risk averse behaviour. Furthermore, the absolute difference in the risk neutral and real world expectations decreases for interest rates with longer maturities. This results from the less variation of interest rates with longer maturities, which is an implicit model characteristic of the Gauss2++ model and is supported by historical data as well. A risk premium is therefore higher (less negative) for a risk averse and lower (less positive) for a risk seeking investor in an arbitrage free market.
\\
\\
Figure \ref{fig:ThreeRPFunctions} shows the absolute risk premium functions of the short-rate for all three modelling types. 
\noindent It can be observed that for the constant and the step function the absolute risk premium is the same up to year $2$. After that year the Gauss2++ variant with the step function has a kink in the absolute risk premium as the local long run risk premium changes to a different level, while the modelling case with the constant function continuous to apporach the long term risk premium determined by the short term interest rate forecasts. The modelling case with the linear function results in a different risk premium for the first $2$ years, but approaches -- without a kink -- the same long term risk premium as the step function. All three functions intersect after $2$ years as this is the forecasting horizon of the short term interest rate forecasts, which were used for the calibration. The absolute risk premium at this time point must be the same for all modelling cases such that the expected interested rates of the model coincide with the forecasts.

\subsection{Backtest}
\label{sec:Backtest}
\begin{figure}[H]%
  \centering
  \subfloat[][]{\includegraphics[width=0.3\linewidth]{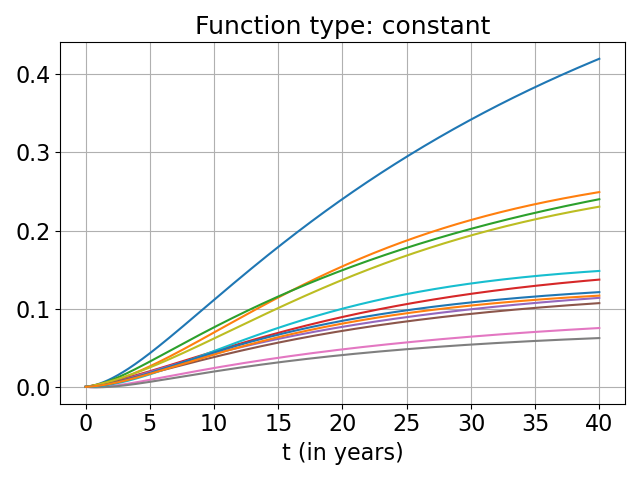}}%
  \subfloat[][]{\includegraphics[width=0.3\linewidth]{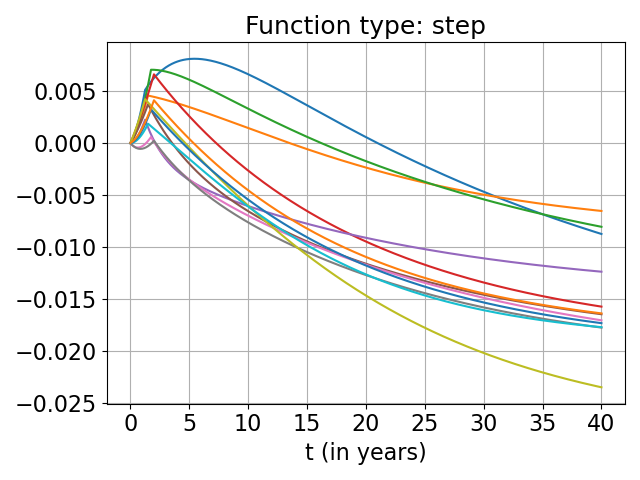}}%
  \subfloat[][]{\includegraphics[width=0.3\linewidth]{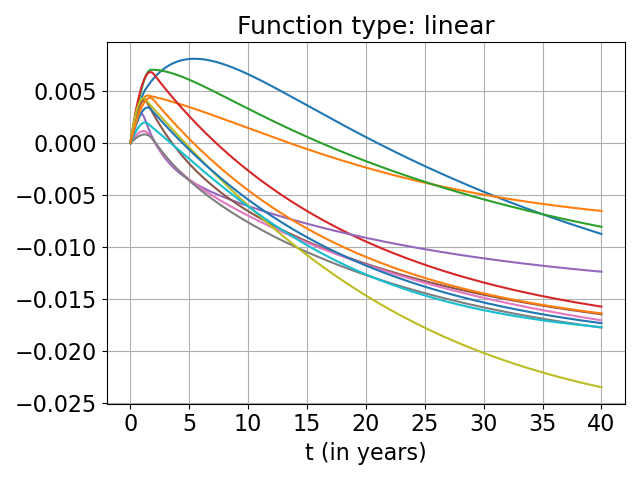}}%
  \subfloat{\includegraphics[width=0.08\linewidth]{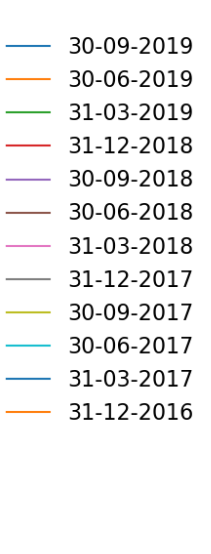}}%
  \caption{Absolute risk premium functions}%
  \label{fig:TotalRPFkt}
\end{figure}

\begin{figure}[H]%
  \centering
  \subfloat[][]{\includegraphics[width=0.3\linewidth]{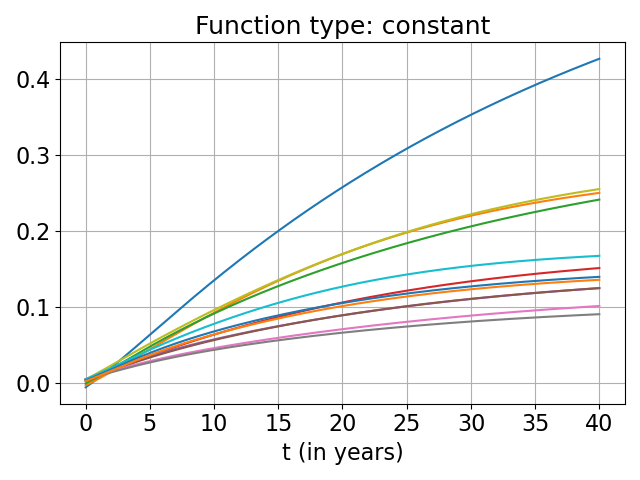}}%
  \subfloat[][]{\includegraphics[width=0.3\linewidth]{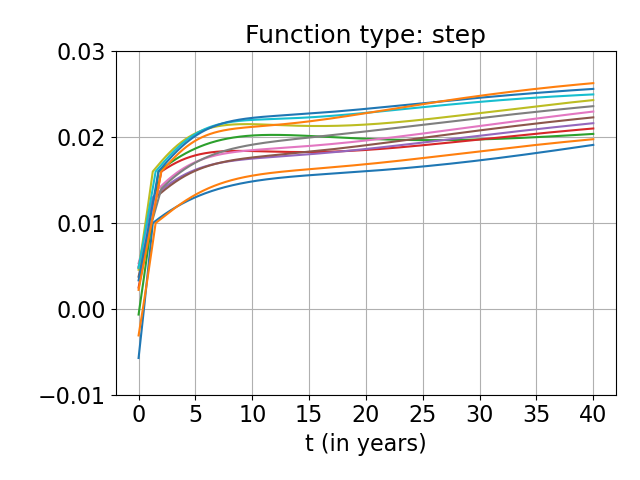}}%
  \subfloat[][]{\includegraphics[width=0.3\linewidth]{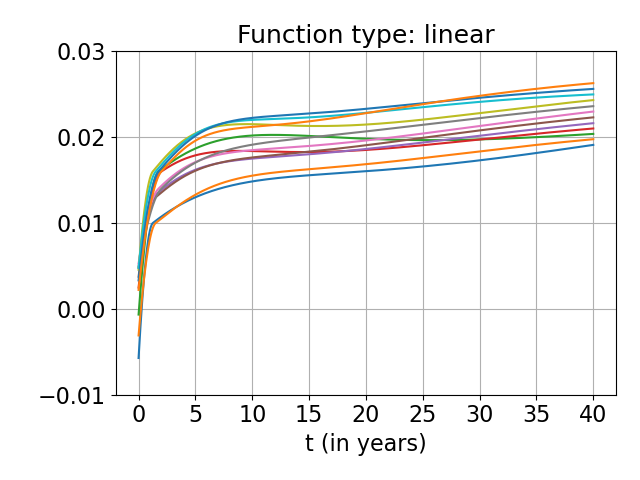}}%
  \subfloat{\includegraphics[width=0.08\linewidth]{Legende.png}}%
  \caption{Development of the expectation of the 10-year interest rate over the modelling horizon for all three variants of the Gauss2++ model}%
  \label{fig:TenYearIRBT}
\end{figure}

In this Subsection the different variants of the Gauss2++ model calibrated on a quarterly basis over the last $3$ years are compared.
\\
\\
As in Section \ref{sec:OneValuationDate} interest rate swaps and swaption volatilities have been used for the risk neutral calibration of the Gauss2++ model. To calibrate the parameters of the local long run risk premium functions in the second calibration step short term interest rate forecasts published by the OECD and a long term average have been used. The forecasts are shown in table (\ref{tbl:IRForecastsOECD}).
\begin{table}
\newcolumntype{M}[1]{>{\centering\arraybackslash}b{#1}}%
\begin{center}
\begin{tabularx}{\textwidth}{M{2cm}M{3cm}M{1.2cm} M{1.2cm} | M{1.2cm} M{1.2cm}}
\textbf{Date}& \multicolumn{3}{c|}{\textbf{Short Term Interest Rate Forecasts}} & \multicolumn{2}{c}{\textbf{Historical Average}}  \\
		    & Forecasting Horizon   & \multicolumn{1}{c}{$3$-m IR} & \multicolumn{1}{c|}{$10$-y IR} & \multicolumn{1}{c}{$3$-m IR} & \multicolumn{1}{c}{$10$-y IR} \\
		    & (in months)& \multicolumn{1}{c}{(in \%)} & \multicolumn{1}{c|}{(in \%)} & \multicolumn{1}{c}{(in \%)} & \multicolumn{1}{c}{(in \%)}\\ \hline \hline
30.09.2019 & $ 15 $ &  $-0.3$  &  $1.0$  &  $1.13$  &  $1.91$\\
30.06.2019 & $ 18 $ &  $-0.3$  &  $1.0$  &  $1.18$  &  $1.98$\\
31.03.2019 & $ 21 $ &  $\textcolor{white}{-}0.2$  &  $1.6$  &  $1.22$  &  $2.04$\\
31.12.2018 & $ 24 $ &  $\textcolor{white}{-}0.2$  &  $1.6$  &  $1.26$  &  $2.10$\\
30.09.2018 & $ 15 $ &  $-0.2$  &  $1.3$  &  $1.31$  &  $2.16$\\
30.06.2018 & $ 18 $ &  $-0.2$  &  $1.3$  &  $1.35$  &  $2.23$\\
31.03.2018 & $ 21 $ &  $-0.3$  &  $1.4$  &  $1.39$  &  $2.30$\\
31.12.2017 & $ 24 $ &  $-0.3$  &  $1.4$  &  $1.44$  &  $2.36$\\
30.09.2017 & $ 15 $ &  $-0.3$  &  $1.6$  &  $1.48$  &  $2.43$\\
30.06.2017 & $ 18 $ &  $-0.3$  &  $1.6$  &  $1.52$  &  $2.50$\\
31.03.2017 & $ 21 $ &  $-0.3$  &  $1.6$  &  $1.57$  &  $2.56$\\
31.12.2016 & $ 24 $ &  $-0.3$  &  $1.6$  &  $1.63$  &  $2.63$\\
\hline
\end{tabularx}
\caption[]{Interest rate forecasts of the OECD and historical average of the 3-month and the 10-year interest rate \footnotemark}
\label{tbl:IRForecastsOECD}
\end{center}
\end{table}
\footnotetext{https://stats.oecd.org}
\noindent The calibration results of the parameters of the Gauss2++ model under the risk neutral measure and of the local long run risk premium function for each variant of the Gauss2++ model can be found in table (\ref{tbl:RNCalibrationResults})-(\ref{tbl:BacktestLinearParam}) in \ref{tbls:BacktestResults}.
\\
\\
For each calibration the absolut risk premium function of the short-rate and the development of the expected 10-year interest rate have been calculated and visualised in Figure \ref{fig:TotalRPFkt} and \ref{fig:TenYearIRBT}.
\noindent The absolute risk premium function of the short-rate for the Gauss2++ model, which uses the constant function for the local long run risk premium, depends highly on the risk neutral calibration results and the forecasts of the OECD. An unfavorable combination of market data and interest rate forecasts can lead to a high value for the local long run risk premium. This value might be reasonable to meet the short term forecasts used for the calibration, but as it stays constant over time it is the value the absolute risk premium is converging to. Therefore, this problem can strike through if the modelling horizon is much longer than the forecasting horizon of the interest rates used for the calibration. In this case a time-varying local long run risk premium function, which can be calibrated to a short and a long term forecast, is more convenient to regularize the risk premium. As it can be seen in Figure \ref{fig:TotalRPFkt} the variants of the Gauss2++ model, which use the step or the linear function for the local long run risk premium, produce more stable risk premiums in the long horizon. In each calibration the absolute risk premium is positive in the first years, which presumes a risk seeking behaviour of the investors, but in the long horizon the absolute risk premium lies between $-0.5\%$ and $-2.5\%$ representing a risk averse market. Also the interest rate distribution in the long horizon is more stable. Figure \ref{fig:TenYearIRBT} (b) and (c) show that the expectation of the 10-year interest rate in the long horizon change only little in each calibration according to the historical average, which was used for the long term interest rate forecast.

\section{Conclusion}
As the Gauss2++ model is often used for pricing purposes, the focus in the literature lies on the evolution of interest rates under the risk neutral measure $\mathbb{Q}$. But regarding risk management and forecasting applications the model under the real world measure is needed. In this paper we introduced a framework to apply the model under both measures in a consistent manner. This framework first conducts a calibration under the risk neutral measure and then determines the change of measure such that it is possible to switch between the risk neutral and the real world. We showed that according to Girsanov this change of measure can be specified by any progressive and square-integrable function without loosing the analytic tractability for e.g. zero-coupon bond prices. 
\citet{hull2014} argue that because of unobserved risk factors, which are not included in the model, a time-varying function should be used, because otherwise unrealistic interest rates in the long forecasting horizon could be reached. We therefore compared a variant of our framework, which uses constant functions to model the change of measure, with two variants, which use either a step or a linear functions. These functions are the simplest extensions of the constant function to a time varying function without increasing the computational effort much. By accounting for different risk premiums in the short and in the long horizon the time varying functions result in much more stable interest rate forecasts in the long run if calibrated at different valuation dates. From a macroeconomical point of view it makes sense that current market fluctuations should not influence interest rate forecasts in the long horizon, e.g. in $40$ years, much. This would also imply that risk measures calculated with the Gauss2++ model, which uses one of the time-varying functions for the change of measure, would be more consistent if estimated at different valuation time points.

\section*{Acknowledgement}
This research was supported by \textbf{ROKOCO predictive analytics GmbH}. We thank our colleagues from \textbf{ROKOCO predictive analytics GmbH} who provided insight and expertise that greatly assisted the research.

\newpage
\bibliographystyle{plainnat}
\bibliography{Literatur}

\begin{thebibliography}{15}
\providecommand{\natexlab}[1]{#1}
\providecommand{\url}[1]{\texttt{#1}}
\expandafter\ifx\csname urlstyle\endcsname\relax
  \providecommand{\doi}[1]{doi: #1}\else
  \providecommand{\doi}{doi: \begingroup \urlstyle{rm}\Url}\fi

\bibitem[Ahmad and Wilmott(2006)]{ahmad2006}
Riaz Ahmad and Paul Wilmott.
\newblock The market price of interest-rate risk: Measuring and modelling fear
  and greed in the fixed-income markets.
\newblock \emph{Wilmott magazine}, pages 64--70, 2006.

\bibitem[Brigo and Mercurio(2007)]{brigo2007}
Damiano Brigo and Fabio Mercurio.
\newblock \emph{Interest rate models -- theory and practice: with smile,
  inflation and credit}.
\newblock Springer Science \& Business Media, 2007.

\bibitem[Cox et~al.(1985)Cox, Ingersoll~Jr, and Ross]{cox1985}
John~C Cox, Jonathan~E Ingersoll~Jr, and Stephen~A Ross.
\newblock An intertemporal general equilibrium model of asset prices.
\newblock \emph{Econometrica: Journal of the Econometric Society}, pages
  363--384, 1985.

\bibitem[Cox and Pedersen(1999)]{cox1999}
Samuel~H Cox and Hal~W Pedersen.
\newblock Nonparameteric estimation of interest rate term structure and
  insurance applications.
\newblock In \emph{Proceedings of the 1999 ASTIN Colloquium, Tokyo, Japan (to
  appear)}, 1999.

\bibitem[Dai and Singleton(2000)]{dai2000}
Qiang Dai and Kenneth~J Singleton.
\newblock Specification analysis of affine term structure models.
\newblock \emph{The journal of finance}, 55\penalty0 (5):\penalty0 1943--1978,
  2000.

\bibitem[Diebold and Li(2006)]{diebold2006}
Francis~X Diebold and Canlin Li.
\newblock Forecasting the term structure of government bond yields.
\newblock \emph{Journal of econometrics}, 130\penalty0 (2):\penalty0 337--364,
  2006.

\bibitem[Duffee(2002)]{duffee2002}
Gregory~R Duffee.
\newblock Term premia and interest rate forecasts in affine models.
\newblock \emph{The Journal of Finance}, 57\penalty0 (1):\penalty0 405--443,
  2002.

\bibitem[Duffie and Kan(1996)]{duffie1996}
Darrell Duffie and Rui Kan.
\newblock A yield-factor model of interest rates.
\newblock \emph{Mathematical finance}, 6\penalty0 (4):\penalty0 379--406, 1996.

\bibitem[Girsanov(1960)]{girsanov1960}
Igor~Vladimirovich Girsanov.
\newblock On transforming a certain class of stochastic processes by absolutely
  continuous substitution of measures.
\newblock \emph{Theory of Probability \& Its Applications}, 5\penalty0
  (3):\penalty0 285--301, 1960.

\bibitem[Heath et~al.(1992)Heath, Jarrow, and Morton]{heath1992}
David Heath, Robert Jarrow, and Andrew Morton.
\newblock Bond pricing and the term structure of interest rates: A new
  methodology for contingent claims valuation.
\newblock \emph{Econometrica: Journal of the Econometric Society}, pages
  77--105, 1992.

\bibitem[Hull and White(1990)]{hullwhite1990}
John Hull and Alan White.
\newblock Pricing interest rate derivative securities.
\newblock \emph{The review of financial studies}, 3\penalty0 (4):\penalty0
  573--592, 1990.

\bibitem[Hull et~al.(2014)Hull, Sokol, and White]{hull2014}
John Hull, Alexander Sokol, and Alan White.
\newblock Short rate joint measure models.
\newblock \emph{Risk}, 10:\penalty0 59--63, 2014.

\bibitem[Jong(2000)]{jong2000}
Frank~de Jong.
\newblock Time series and cross-section information in affine term-structure
  models.
\newblock \emph{Journal of Business \& Economic Statistics}, 18\penalty0
  (3):\penalty0 300--314, 2000.

\bibitem[Stanton(1997)]{stanton1997}
Richard Stanton.
\newblock A nonparametric model of term structure dynamics and the market price
  of interest rate risk.
\newblock \emph{The Journal of Finance}, 52\penalty0 (5):\penalty0 1973--2002,
  1997.

\bibitem[Vasicek(1977)]{vasicek1977}
Oldrich Vasicek.
\newblock An equilibrium characterization of the term structure.
\newblock \emph{Journal of financial economics}, 5\penalty0 (2):\penalty0
  177--188, 1977.

\end{thebibliography}

\appendix


\section{Bond Price Dynamic under the Risk Neutral Measure}
\label{proof:BPdynamicQ}
By defining
\begin{align*}
A(t,T) &= -\int_t^T \varphi(s) ds + \frac{1}{2}V(t,T),
\end{align*}
the price of a zero-coupon bond $P(t,T)$ at time point $t$ and maturity $T$ can be calculated for the Gauss2++ model under the risk neutral measure $\mathbb{Q}$ by
\begin{align}
\label{eq:BPFormula2}
P(t,T) = e^{A(t,T) - B(a,t,T) x(t)-B(b,t,T)y(t)}.
\end{align}
A proof of this formula can be found in \citep{brigo2007}. The derivatives of $A(t,T)$ and $V(t,T)$ with respect to the first entry and of $B(z,t,T)$ with respect to the second entry are given by
\begin{align*}
A'(t,T) &= \varphi(t) + \frac{1}{2} V'(t,T), \\
V'(t,T) &= -\sigma^2B(a,t,T)^2 - \eta^2 B(b,t,T)^2 - 2 \sigma \eta \rho B(a,t,T) B(b,t,T), \\
B'(z, t, T) &= -e^{-z(T-t)}.
\end{align*}
Furthermore, it holds
$$
B(z,t,T)z - B'(z,t,T) = 1.
$$
To calculate the zero-coupon bond price dynamic, we apply Itô's formula to (\ref{eq:BPFormula2}), i.e., \par
\begin{scriptsize}
\begin{alignat*}{2}
  dP(t,T) &=  P(t,T) \left[A(t,T)-B(a,t,T)x(t)-B(b,t,T)y(t)\right]'dt &&+ P(t,T)(-B(a,t,T))dx(t) \\
   &{} &&+ P(t,T)(-B(b,t,T))dy(t) \\
   &{} &&+ \frac{1}{2}P(t,T)B(a,t,T)^2\sigma^2dt \\
   &{} &&+ \frac{1}{2}P(t,T)B(b,t,T)^2\eta^2dt \\
   &{} &&+ P(t,T)B(a,t,T)B(b,t,T)\sigma \eta \rho dt \\
   &=P(t,T)\bigg[A'(t,T) - B'(a,t,T)x(t)\hspace{0.1cm}-\hspace{0.1cm} B'(b,t,T)y(t) &&+ B(a,t,T)ax(t) + B(b,t,T)by(t) \\
   &{} &&+\frac{1}{2}B(a,t,T)^2\sigma^2 + \frac{1}{2} B(b,t,T)^2\eta^2 \\
   &{} &&+B(a,t,T)B(b,t,T)\sigma\eta\rho \bigg] dt \\
   &{} &&-B(a,t,T)P(t,T)\sigma dW^1(t) \\
   &{} &&-B(b,t,T)P(t,T)\eta dW^2(t) \\
   &=P(t,T)[\varphi(t) +x(t)+y(t)]dt-B(a,t,T)P(t,T)\sigma dW^1&&(t)-B(b,t,T)P(t,T)\eta dW^2(t) \\
   &=P(t,T)r(t)dt-B(a,t,T)P(t,T)\sigma dW^1(t)-B(b,t,T)&&P(t,T)\eta dW^2(t).
\end{alignat*}
\end{scriptsize}\par

\section{The Dynamics of the Gauss2++ Factors $\boldsymbol{x}$ and $\boldsymbol{y}$ under the Real World Measure}
\label{proof:xyUnderP}
The dynamics of the two processes $x$ and $y$ under the risk neutral measure $\mathbb{Q}$ can be expressed in terms of two independent Brownian motions $\widehat{W}^1$ and $\widehat{W}^2$, i.e.
\begin{align*}
dx(t) &= -ax(t)dt + \sigma d\widehat{W}^1(t), \\
dy(t) &= -by(t)dt + \eta \rho d\widehat{W}^1(t) + \eta \sqrt{(1-\rho^2)}d\widehat{W}^2(t),
\end{align*}
where
\begin{align*}
dW^1(t) &= d\widehat{W}^1(t), \\
dW^2(t) &= \rho d\widehat{W}^1(t) + \sqrt{(1-\rho^2)}d\widehat{W}^2(t).
\end{align*}
According to Girsanov's theorem , as $\boldsymbol{\widehat{W}} = (\widehat{W}^1, \widehat{W}^2)$ is a standard 2-dimensional Brownian motion and let $(\boldsymbol{\Phi}(t))_{t\in[0,\mathcal{T}]} = (\Phi^1(t), \Phi^2(t))_{t\in[0,\mathcal{T}]}$ be a progressive and square-integrable process, the process $\boldsymbol{\breve{W}}$ defined by
\begin{align*}
\boldsymbol{\breve{W}}(t) := \boldsymbol{\widehat{W}}(t) + \int_0^t\boldsymbol{\Phi}(s) ds
\end{align*}
is a standard 2-dimensional Brownian motion under a new measure, which we call $\mathbb{P}$ and declare to be the real world measure. This means that the dynamic of the two Brownian motion $\widehat{W}^1$ and $\widehat{W}^2$ under the real world measure $\mathbb{P}$ is given by
\begin{align*}
d\widehat{W}^1(t) =  d\breve{W}^1(t) - \Phi^1(t)dt, \\
d\widehat{W}^2(t) =  d\breve{W}^2(t) - \Phi^2(t)dt.
\end{align*}
Therefore, the dynamics of the two processes $x$ and $y$ under the real world measure are then given by
\begin{alignat*}{2}
dx(t) &= \bigg[-\Phi^1(t)\sigma - ax(t)\bigg]dt + \sigma d\breve{W}^1(t), \\
dy(t) &= \bigg[-\Phi^1(t)\eta \rho - \Phi^2(t)\eta \sqrt{(1-\rho^2)}-by(t)\bigg]dt &&+ \eta \rho d\breve{W}^1(t) \\
&{} &&+ \eta \sqrt{(1-\rho^2)}d\breve{W}^2(t).
\intertext{\normalsize If we specify $\Phi(t)$ as in (\ref{eq:PhiDefinition}) this simplifies to}
dx(t) &= a(d_x(t) - x(t))dt + \sigma d\breve{W}^1(t), \\
dy(t) &= b(d_y(t) - y(t))dt + \eta \rho d\breve{W}^1(t) + \eta \sqrt{(1-\rho^2)}&&d\breve{W}^2(t).
\end{alignat*}
Representing the dynamics by two correlated Brownian motions $\widetilde{W}^1$ and $\widetilde{W}^2$ results in the equations given in (\ref{eq:xUnderP}) and (\ref{eq:yUnderP}).

\section{Bond Price Dynamic under the Real World Measure}
\label{proof:BPdynamicP}
The dynamic of a zero-coupon bond price $P(t,T)$ under the risk neutral measure $\mathbb{Q}$ expressed by the two independent Brownian motions 
$\widehat{W}^1$ and $\widehat{W}^2$ is given by 
\begin{alignat*}{2}
dP(t,T) &= P(t,T)r(t)dt &&- P(t,T)B_{\tau}(a)\sigma d\widehat{W}^1(t)
- P(t,T)B_{\tau}(b) \eta \rho d\widehat{W}^1(t) \\
&{} &&- P(t,T)B_{\tau}(b)\eta \sqrt{(1-\rho^2)}d\widehat{W}^2(t), \\
&= P(t,T)r(t)dt &&- \bigg[P(t,T)B_{\tau}(a)\sigma + P(t,T)B_{\tau}(b) \eta \rho \bigg] d\widehat{W}^1(t)\\
&{} &&- P(t,T)B_{\tau}(b)\eta \sqrt{(1-\rho^2)}d\widehat{W}^2(t). \\
\end{alignat*}
Applying Girsanov's theorem as in appendix \ref{proof:xyUnderP} the dynamic under the real world measure $\mathbb{P}$ amounts to \par
\begin{scriptsize}
\begin{alignat*}{2}
dP(t,T) &= P(t,T)r(t)dt &&- \bigg[P(t,T)B_{\tau}(a)\sigma + P(t,T)B_{\tau}(b) \eta \rho \bigg] d\widehat{W}^1(t) \\
&{} &&- P(t,T)B_{\tau}(b)\eta \sqrt{(1-\rho^2)}d\widehat{W}^2(t) \\
&= P(t,T)\bigg[r(t) &&+ \bigg(B_{\tau}(a) \sigma + B_{\tau}(b) \eta \rho \bigg) \left(-\frac{a d_x(t)}{\sigma}\right)\\
&{} &&+ B_{\tau}(b) \eta \sqrt{(1-\rho^2)} \left(-\frac{bd_y(t)}{\eta \sqrt{(1-\rho^2)}}+\frac{\rho a d_x(t)}{\sigma \sqrt{(1-\rho^2)}}\right)\bigg] dt \\
&{} &&- \bigg[P(t,T)B_{\tau}(a)\sigma + P(t,T)B_{\tau}(b) \eta \rho \bigg]  d\breve{W}^1(t) \\
&{} &&- P(t,T)B_{\tau}(b)\eta \sqrt{(1-\rho^2)}d\breve{W}^2(t) \\
&= P(t,T)\bigg[r(t) &&- B_{\tau}(a) a d_x(t) - B_{\tau}(b) b d_y(t)\bigg] dt \\
&{} &&- \bigg[P(t,T)B_{\tau}(a)\sigma + P(t,T)B_{\tau}(b) \eta \rho \bigg]  d\breve{W}^1(t) \\
&{} &&- P(t,T)B_{\tau}(b)\eta \sqrt{(1-\rho^2)}d\breve{W}^2(t).
\end{alignat*}
\end{scriptsize} \par
\noindent Representing the dynamic by two correlated Brownian motions $\widetilde{W}^1$ and $\widetilde{W}^2$ results in the equation given in (\ref{eq:BPdynamicRW}).

\section{Individual Discount Rate for the Zero-Coupon Bonds in the Real World}
\label{proof:DiskRateP}
\begin{proof}
To proof that $\frac{P(t,T)}{X(t,T)}$ is indeed a martingale we calculate the dynamic of the discounted price process.
\begin{footnotesize}
\begin{alignat*}{2}
d\frac{P(t,T)}{X(t)} &= d\big(\frac{1}{X(t)} \cdot P(t&&,T) \big) \\
&= \frac{1}{X(t)}dP(t,T&&) + P(t,T)d\frac{1}{X(t)} + d\left< P(t,T), \frac{1}{X(t)} \right > \\
&= \frac{1}{X(t)}dP(t,T&&)- \frac{P(t,T)}{X(t)}\left[r(t) - B(a,t,T)ad_x(t) - B(b,t,T)bd_y(t)\right]dt \\
&= \frac{P(t,T)}{X(t)} \big[r(t)&&-B(a,t,T)ad_x(t) - B(b,t,T)bd_y(t)\big]dt \\
&{} &&- \frac{P(t,T)}{X(t)} B(a,t,T)\sigma d\widetilde{W}^1(t)-\frac{P(t,T)}{X(t)} B(b,t,T)\eta d\widetilde{W}^2(t)\\
&{} &&- \frac{P(t,T)}{X(t)}\left[r(t) - B(a,t,T)ad_x(t) - B(b,t,T)bd_y(t)\right]dt \\
&= -\frac{P(t,T)}{X(t)}B(&&a,t,T)\sigma d\widetilde{W}^1(t)- \frac{P(t,T)}{X(t)} B(b,t,T)\eta d\widetilde{W}^2(t)
\end{alignat*}
\end{footnotesize}
\end{proof}

\section{Bond Price Formula under the Real World Measure}
\label{proof:BPFormulaP}
To calculate the price of a zero-coupon bond under the real world measure $\mathbb{P}$, the distribution of 
\begin{align*}
exp\left(-\int_t^T \left(r(u) - B(a,u,T) a d_x(u) - B(b,u,T) bd_y(u) \right) du \right)
\end{align*}
has to be determined. In the following we show, that the integral in the exponent is normaly distributed and calculate the mean and the variance of
\begin{align}
\label{eq:SRIntegral}
I(t,T) \coloneqq \int_t^T \left(r(u) - B(a,u,T) a d_x(u) - B(b,u,T) bd_y(u) \right) du.
\end{align}
We first concentrate on the integral over the short-rate $r(s)$, which is a sum of the $x$- and the $y$-process and a deterministic function
$$
r(s) = x(s) + y(s) + \varphi(s).
$$
The integral over the process $x$ is given by 
\begin{alignat*}{2}
\int_t^T x(u) du &=  \int_t^T  \bigg(x(t) e^{-a(u-t)} &&+ \int_t^u ae^{-a(u-s)}d_x(s)ds \\
&{} &&+ \int_t^u \sigma e^{-a(u-s)}d\widetilde{W}^1(s)\bigg)du \\
&= \underbrace{\int_t^T  x(t) e^{-a(u-t)} du}_{\larger\textcircled{\smaller[2]1}} &&+ \underbrace{\int_t^T \int_t^u ae^{-a(u-s)}d_x(s)dsdu}_{\larger\textcircled{\smaller[2]2}}\\
&{} &&+ \underbrace{\int_t^T\int_t^u \sigma e^{-a(u-s)}d\widetilde{W}^1(s)du.}_{\larger\textcircled{\smaller[2]3}}
\end{alignat*}
The first integral amounts to
\begin{align*}
{\larger\textcircled{\smaller[2]1}} &= x(t) \int_t^Te^{-a(u-t)}du = x(t)\left[-\frac{1}{a}e^{-a(u-t)} \right]_t^T = x(t) \frac{1-e^{-a(T-t)}}{a}. \\
\intertext{For the second integral we use the integration by parts formula}
{\larger\textcircled{\smaller[2]2}} &= \int_t^T\left(\int_t^u e^{as}d_x(s)ds\right)ae^{-au}du \\
&= a \int_t^T\left(\int_t^u e^{as}d_x(s)ds\right)d_u\left(\int_t^ue^{-av}dv\right) \\
&= a \left[ \left( \int_t^Te^{au}d_x(u)du\right)\left(\int_t^Te^{-av}dv\right) - \int_t^T\left(\int_t^ue^{-av}dv\right)e^{au}d_x(u)du\right]\\
&= a\left[\int_t^T\left(\int_u^Te^{-av}dv\right)e^{au}d_x(u)du \right] \\
&= \int_t^T\left(1-e^{-a(T-u)}\right)d_x(u)du \\
&= \int_t^TaB(a,u,T)d_x(u)du. \\
\intertext{For the third integral we again use the integration by parts formula}
{\larger\textcircled{\smaller[2]3}} &= \sigma \int_t^T\left(\int_t^u e^{as}d\widetilde{W}^1(s)\right)ae^{-au}du \\
&= \sigma \int_t^T\left(\int_t^u e^{as}d\widetilde{W}^1(s)\right)d_u\left(\int_t^ue^{-av}dv\right) \\
&= \sigma \left[ \left( \int_t^Te^{au}d\widetilde{W}^1(u)\right)\left(\int_t^Te^{-av}dv\right) - \int_t^T\left(\int_t^ue^{-av}dv\right)e^{au}d\widetilde{W}^1(u)\right]\\
&= \sigma\left[\int_t^T\left(\int_u^Te^{-av}dv\right)e^{au}d\widetilde{W}^1(u) \right] \\
&= \sigma \int_t^T\left[-\frac{e^{-av}}{a} \right]_u^T e^{au}d\widetilde{W}^1(u) \\
&= \frac{\sigma}{a} \int_t^T\left(1-e^{-a(T-u)}\right)d\widetilde{W}^1(u) \\
&= \frac{\sigma}{a}\int_t^T \left(1-e^{-a(T-u)}\right)d\widetilde{W}^1(u).
\end{align*}
The corresponding expressions for $\int_t^T y(u) du$ can be obtained analogously. We observe that the results of integral ${\larger\textcircled{\smaller[2]2}}$ for $\int_t^T x(u)du$ and $\int_t^T y(u)du$ cancel out with the last two terms in equation (\ref{eq:SRIntegral}). Therefore it remains
\begin{align*}
I(t,T) &= \int_t^T \varphi(u)du + \frac{1-e^{-a(T-t)}}{a} x(t) + \frac{1-e^{-b(T-t)}}{b} y(t) \\
&+ \frac{\sigma}{a}\int_t^T \left(1-e^{-a(T-u)}\right)d\widetilde{W}^1(u) + \frac{\eta}{b}\int_t^T \left(1-e^{-b(T-u)}\right)d\widetilde{W}^2(u).
\end{align*}
As $\boldsymbol{\widetilde{W}} = (\widetilde{W}^1, \widetilde{W}^2)$ is a 2-dimensional Brownian motion under $\mathbb{P}$, $I(t,T)$ is normally distributed and the mean and the variance can be easily retrieved resulting in (\ref{eq:RWmean}) and (\ref{eq:RWvariance}).

\section{Tables of Backtest Results}
\label{tbls:BacktestResults}
\begin{table}[H]
\newcolumntype{M}[1]{>{\centering\arraybackslash}b{#1}}%
\begin{center}
\begin{tabularx}{\textwidth}{M{2.2cm}M{1.5cm}M{1.5cm}M{1.5cm}M{1.5cm}M{1.5cm}}
	Date   & $a$    &  $b$   & $\sigma$ & $\eta$  & $\rho$ \\ \hline \hline
30.09.2019 & $ 0.2694 $ & $ 0.0269 $ & $ 0.0121 $ & $ 0.0089 $ & $ -0.8950 $ \\ 
30.06.2019 & $ 0.1216 $ & $ 0.0628 $ & $ 0.0363 $ & $ 0.0283 $ & $ -0.9687 $ \\ 
31.03.2019 & $ 0.3978 $ & $ 0.0331 $ & $ 0.0333 $ & $ 0.0091 $ & $ -0.8576 $ \\ 
31.12.2018 & $ 0.1628 $ & $ 0.0521 $ & $ 0.0183 $ & $ 0.0154 $ & $ -0.8629 $ \\ 
30.09.2018 & $ 0.6100 $ & $ 0.0429 $ & $ 0.0459 $ & $ 0.0104 $ & $ -0.8722 $ \\ 
30.06.2018 & $ 0.2901 $ & $ 0.0459 $ & $ 0.0104 $ & $ 0.0112 $ & $ -0.9941 $ \\ 
31.03.2018 & $ 0.5120 $ & $ 0.0386 $ & $ 0.0142 $ & $ 0.0097 $ & $ -1.0000 $ \\ 
31.12.2017 & $ 0.3803 $ & $ 0.0471 $ & $ 0.0236 $ & $ 0.0120 $ & $ -0.8854 $ \\ 
30.09.2017 & $ 0.0880 $ & $ 0.0655 $ & $ 0.0421 $ & $ 0.0460 $ & $ -0.9938 $ \\
30.06.2017 & $ 0.1260 $ & $ 0.0890 $ & $ 0.0504 $ & $ 0.0517 $ & $ -0.9963 $ \\
31.03.2017 & $ 0.2940 $ & $ 0.0581 $ & $ 0.0152 $ & $ 0.0146 $ & $ -0.9984 $ \\
31.12.2016 & $ 0.2427 $ & $ 0.0606 $ & $ 0.0178 $ & $ 0.0173 $ & $ -1.0000 $ \\ \hline
\end{tabularx}
\caption{Calibration results of the risk neutral calibration on a quarterly basis from 31.12.2016 until 30.09.2019}
\label{tbl:RNCalibrationResults}
\end{center}
\end{table}
\begin{table}[H]
\begin{center}
\newcolumntype{M}[1]{>{\centering\arraybackslash}b{#1}}%
\begin{tabularx}{\textwidth}{M{2.2cm}M{4.5cm}M{4.5cm}}
	Date    & $d_x$    & $d_y$   \\ \hline \hline
30.09.2019  & $-0.0676$  &  $0.7400$ \\
30.06.2019  & $-0.2848$  &  $0.5787$ \\ 
31.03.2019  & $-0.0267$  &  $0.3636$ \\ 
31.12.2018  & $-0.0539$  &  $0.2182$ \\ 
30.09.2018  & $-0.0107$  &  $0.1518$ \\ 
30.06.2018  & $-0.0173$  &  $0.1481$ \\ 
31.03.2018  & $-0.0112$  &  $0.1099$ \\ 
31.12.2017  & $-0.0150$  &  $0.0913$ \\ 
30.09.2017  & $-0.7023$  &  $0.9836$ \\ 
30.06.2017  & $-0.3883$  &  $0.5497$ \\
31.03.2017	& $-0.0330$  &  $0.1710$ \\ 
31.12.2016  & $-0.0405$  &  $0.1725$ \\ 
\hline
\end{tabularx}
\caption{Quarterly calibration results for the constant local long run risk premium functions from 31.12.2016 to 30.09.2019.}
\label{tbl:BacktestConstantParam}
\end{center}
\end{table}
\begin{table}[H]
\begin{center}
\newcolumntype{M}[1]{>{\centering\arraybackslash}b{#1}}%
\begin{tabularx}{\textwidth}{M{2.2cm}M{2cm}M{2cm}M{2cm}M{2cm}}
	Date    & $d_x$    & $d_y$    & $l_x$	  & $l_y$    \\ \hline \hline
30.09.2019  &  $-0.0676$  &  $0.7400$  & $-0.0090$  & $-0.0129$  \\ 
30.06.2019  &  $-0.2848$  &  $0.5787$  & $-0.0376$  & $\textcolor{white}{-}0.0292$  \\ 
31.03.2019  &  $-0.0267$  &  $0.3636$  & $-0.0114$  & $-0.0034$  \\ 
31.12.2018  &  $-0.0539$  &  $0.2182$  & $-0.0163$  & $-0.0029$  \\ 
30.09.2018  &  $-0.0107$  &  $0.1518$  & $-0.0101$  & $-0.0047$  \\ 
30.06.2018  &  $-0.0173$  &  $0.1481$  & $-0.0107$  & $-0.0090$  \\ 
31.03.2018  &  $-0.0112$  &  $0.1099$  & $-0.0087$  & $-0.0129$  \\ 
31.12.2017  &  $-0.0150$  &  $0.0913$  & $-0.0099$  & $-0.0111$  \\ 
30.09.2017  &  $-0.7023$  &  $0.9836$  & $-0.0364$  & $\textcolor{white}{-}0.0087$  \\ 
30.06.2017  &  $-0.3883$  &  $0.5497$  & $-0.0423$  & $\textcolor{white}{-}0.0233$  \\ 
31.03.2017  &  $-0.0330$  &  $0.1710$  & $-0.0131$  & $-0.0068$  \\ 
31.12.2016  &  $-0.0405$  &  $0.1725$  & $-0.0154$  & $-0.0033$  \\ \hline
\end{tabularx}
\caption{Quarterly calibration results for the step local long run risk premium functions from 31.12.2016 to 30.09.2019.}
\label{tbl:BacktestStepParam}
\end{center}
\end{table}
\begin{table}[H]
\begin{center}
\newcolumntype{M}[1]{>{\centering\arraybackslash}b{#1}}%
\begin{tabularx}{\textwidth}{M{2.2cm}M{2cm}M{2cm}M{2cm}M{2cm}}
 Date  	    & $d_x$     & $d_y$    & $l_x$    &  $l_y$ \\ \hline \hline
30.09.0219  &  $-0.1332$  &  $1.5015$  & $-0.0090$  & $-0.0129$ \\ 
30.06.2019  &  $-0.5474$  &  $1.1457$  & $-0.0376$  & $\textcolor{white}{-}0.0292$ \\ 
31.03.2019  &  $-0.0461$  &  $0.7377$  & $-0.0114$  & $-0.0034$ \\ 
31.12.2018  &  $-0.0959$  &  $0.4471$  & $-0.0163$  & $-0.0029$ \\ 
30.09.2018  &  $-0.0114$  &  $0.3111$  & $-0.0101$  & $-0.0047$ \\ 
30.06.2018  &  $-0.0250$  &  $0.3087$  & $-0.0107$  & $-0.0090$ \\ 
31.03.2018  &  $-0.0144$  &  $0.2355$  & $-0.0087$  & $-0.0129$ \\ 
31.12.2017  &  $-0.0216$  &  $0.1970$  & $-0.0099$  & $-0.0111$ \\ 
30.09.2017  &  $-1.3930$  &  $1.9854$  & $-0.0364$  & $\textcolor{white}{-}0.0087$ \\ 
30.06.2017  &  $-0.7567$  &  $1.1001$  & $-0.0423$  & $\textcolor{white}{-}0.0233$ \\ 
31.03.2017  &  $-0.0567$  &  $0.3550$  & $-0.0131$  & $-0.0068$ \\ 
31.12.2016  &  $-0.0700$  &  $0.3556$  & $-0.0154$  & $-0.0033$ \\ \hline
\end{tabularx}
\caption{Quarterly calibration results for the linear local long run risk premium functions from 31.12.2016 to 30.09.2019.}
\label{tbl:BacktestLinearParam}
\end{center}
\end{table}

\end{document}